\documentclass[useAMS,usenatbib]{mn2e}
\usepackage{times,mathptm}
\usepackage{lscape}

\usepackage{natbib}
\usepackage{graphicx}
\usepackage{floatflt}
\usepackage{amssymb}

\newcommand{\lam}{$\lambda$}

\newcommand{\sig}{$\sigma$}

\newcommand{\cs}{$\textnormal{cm}^{-2}$}
\newcommand{\mujy}{$\umu$Jy}
\newcommand{\mjy}{mJy}
\newcommand{\ergcms}{erg~s$^{-1}$~cm$^{-2}$}
\newcommand{\ergs}{erg~s$^{-1}$}
\newcommand{\Lsun}{L$_\odot$}
\newcommand{\MJysr}{$\textnormal{MJy~sr}^{-1}$}

\newcommand{\e}[1]{$\times 10^{#1}$}
\newcommand{\fratio}{$S_{70}/S_{24}$}
\newcommand{\lratio}{$L_{\rm IR}/L_{\rm X}$}
\newcommand{\lsixratio}{$\nu L_{\nu}(6\mu m)/L_{\rm X}$}
\newcommand{\vlvratio}{$\nu L_{\nu}(6\mu m)/L_{\rm X}$}

\newcommand{\Tab}[1]{Table \ref{#1}}
\newcommand{\Fig}[1]{Fig. \ref{#1}}

\newcommand{\highz}{high--$z$}

\newcommand{\Nh}{$N_{\rm H}$}
\newcommand{\Lx}{$L_{\rm X}$}
\newcommand{\Lir}{$L_{\rm IR}$}
\newcommand{\comment}[1]{}

\newcommand{\Spitzer}{\textit{Spitzer}}
\newcommand{\Herschel}{\textit{Herschel}}
\newcommand{\IRAS}{\textit{IRAS}}
\newcommand{\Chandra}{\textit{Chandra}}
\newcommand{\XMM}{\textit{XMM-Newton}}
\newcommand{\Swift}{\textit{Swift}}

\title[FIR properties of CDF-S X-ray AGNs]{Characterising the
  Far-infrared Properties of Distant X-ray Detected AGNs: Evidence for Evolution in the Infrared--X-ray
  Luminosity Ratio}
\author[J. R. Mullaney et al.]{J. R. Mullaney$^{1}$\thanks{E-mail: j.r.mullaney@dur.ac.uk}, D. M. Alexander$^{1}$, M. Huynh$^{2}$, A. D. Goulding$^{1}$ \& D. Frayer$^{2}$\\
  $^{1}$Department of Physics and Astronomy, Durham University, South Road, Durham, DH1 3LE, U.K.\\
  $^{2}$Infrared Processing and Analysis Center, California Institute
  of Technology, 100-22, Pasadena, CA 91125}
\begin{document}

\date{Date Accepted}

\pagerange{\pageref{firstpage}--\pageref{lastpage}} \pubyear{2007}

\maketitle

%
\begin{abstract} 
%

  We investigate the far-infrared properties of X-ray sources detected
  in the {\it Chandra} Deep Field-South (CDF-S) survey using the
  ultra-deep 70~\micron\ and 24~\micron\ {\it Spitzer} observations
  taken in this field. Since only 30 (i.e., $\approx10$\%) of the 266
  X-ray sources in the region of the 70~\micron\ observations are
  detected at 70~\micron, we rely on stacking analyses of the
  70~\micron\ data to characterise the average 70~\micron\ properties
  of the X-ray sources as a function of redshift, X-ray luminosity and
  X-ray absorption.  Using \Spitzer -IRS data of the \Swift -BAT
  sample of $z\approx0$ active galactic nuclei (hereafter, AGNs), we
  show that the 70/24~\micron\ flux ratio can distinguish between
  AGN-dominated and starburst-dominated systems out to $z\approx 1.5$.
  Among the X-ray sources detected at 70~\micron\ we note a large
  scatter in the observed 70/24~\micron\ flux ratios, spanning almost
  a factor of 10 at similar redshifts, irrespective of object
  classification, suggesting a range of AGN:starburst ratios.  From
  stacking analyses we find that the average observed 70/24~\micron\
  flux ratios of AGNs out to an average redshift of 1.5 are similar to
  $z\approx0$ AGNs with similar X-ray luminosities
  ($L_{X}=10^{42-44}$~\ergs) and absorbing column densities
  ($N_H\leq10^{23}$~\cs). Furthermore, both high redshift and
  $z\approx0$ AGNs follow the same tendency toward warmer
  70/24~\micron\ colours with increasing X-ray luminosity ($L_{X}$).
  From analyses of the \Swift-BAT sample of $z\approx0$ AGNs, we note
  that the 70~\micron\ flux can be used to determine the infrared
  (8-1000~\micron) luminosities of high redshift AGNs.  We use this
  information to show that \Lx$ = 10^{42-43}$ \ergs AGNs at high
  redshifts ($z=1-2$) have infrared to X-ray luminosity ratios
  (hereafter, \lratio) that are, on average, $4.7_{-2.0}^{+10.2}$ and
  $12.7^{+7.1}_{-2.6}$ times higher than AGNs with similar X-ray
  luminosities at $z=0.5-1$ and $z\approx0$, respectively.  By
  comparison, we find that the \lratio\ ratios of \Lx$ = 10^{43-44}$
  \ergs\ AGNs remain largely unchanged across this same redshift
  interval.  We explore the consequences that these results may have
  on the identification of distant, potentially Compton-thick AGNs
  using \lratio\ ratios.  In addition, we discuss possible scenarios
  to account for the observed increase in the \lratio\ ratio with
  redshift, including changes in the dust covering factor of AGNs
  and/or the star formation rates of their host galaxies.  Finally, we
  show how deep observations to be undertaken by the \textit{Herschel
    Space Observatory} will enable us to discriminate between these
  proposed scenarios and also identify Compton-thick AGNs at high
  redshifts.

\end{abstract}


%
\section{Introduction}
%
\label{Intro}
Deep X-ray surveys undertaken by the \Chandra\ and \XMM\ observatories
have provided the most efficient method to date of identifying large
numbers of active galactic nuclei (hereafter, AGNs) out to high
redshifts ($z\approx5$; e.g., \citealt{Alexander01, Barger03, Bauer04,
  Szokoly04, Brandt05, Mainieri05}). Due to the high penetrating power
of hard (i.e.,\ $>$2~keV) X-rays, they provide a means of identifying
AGNs that is much less biased toward unobscured AGNs than, for
example, that obtained at optical wavelengths alone.  Consequently,
deep X-ray surveys have revealed that a large fraction (up to
$\approx$~75\%; \citealt{Mainieri02, Dwelly06, Tozzi06, Tajer07})  of highly obscured AGNs are often unidentified or misclassified by
observations at other wavelengths (e.g. \citealt{Barger03,
    Szokoly04}). Furthermore, as the central engines of AGNs are the
only compact objects capable of producing luminous X-ray emission
(i.e.,\ $L_{\rm X}>10^{42}$~\ergs), observations at these high
energies provide the most reliable method of measuring the
intrinsic power of the central engine without suffering from
significant contamination by star-formation.

Although X-rays provide an ideal means of studying the central engines
of AGNs, they yield limited information on the regions outside the
central few parsecs. For example, although X-ray observations have
been used extensively to measure the amount of absorbing gas
(i.e. \Nh) \textit{along our line of sight} to the central engine  (e.g. \citealt{Mushozky93, Bassani99, Malizia03, Guainazzi05}),
they tell us little about the spatial extent and covering factor of
the putative obscuring dusty torus, a key component of the
unification models that have become a cornerstone of AGN research over
the past two decades (e.g. \citealt{Antonucci85, Krolik88,
  Antonucci93}). Furthermore, as any X-ray emission from stellar
processes will be swamped by the presence of an AGN, X-rays alone
provide little insight into the relationship between star-formation
and AGN activity that is predicted by both galaxy formation models
(e.g. \citealt{Kauffmann00, Benson03, Granato04, Bower06, Booth09})
and locally defined black hole--bulge relationships
(e.g. \citealt{Magorrian98, Gebhardt00, Tremaine02,
  Mclure02}). Observations at infrared (hereafter, IR) wavelengths can
provide significant insights on both these counts.

Like X-rays, the longer IR wavelengths (i.e., $>$5~\micron) are
largely unaffected by absorption from interstellar gas/dust. However,
because IR radiation is typically produced via the reprocessing of
shorter wavelength light by dust, it also provides a means to study
the dusty environments surrounding AGNs. Furthermore, as both AGNs and
starbursting systems are capable of producing large amounts of IR
radiation, the study of AGNs at these wavelengths can provide insights
into the processes that connects AGN activity and star-formation.
However, despite considerable gains in our understanding of the IR
properties of AGNs since the launch of the \Spitzer\ {\it Space
  Telescope}, there remains many unresolved questions. For example, we
currently lack a clear picture of how the IR emission of AGNs is
affected by the differences in the intrinsic power of the central
engine and, perhaps more importantly, whether it evolves over
cosmic time.  As a result, our understanding of both the putative
torus and the interplay between AGN activity and star-formation
is limited.

To date, most studies of the IR properties of AGNs have
concentrated on the relatively short wavelength data provided by the
IRAC instrument (3.6~\micron, 4.5~\micron, 5.8~\micron\ and
8.0~\micron) and the 24~\micron\ observations taken with the MIPS
instrument on-board the \Spitzer\ \textit{Space Telescope}
(e.g. \citealt{Lutz04,Stern05,Donley07,Daddi07}). However,
studies of high redshift objects at these shorter wavelengths are
hindered by the presence of strong spectral features at \lam\lam
  $\lesssim$ 12 \micron\ (in particular polycyclic aromatic
hydrocarbons [PAHs], silicate absorption features and strong emission
lines) that shift into these wavebands at $z \gtrsim 0.7$. For
example, the IR luminosities of high redshift galaxies are
overestimated by a factor of $\sim 2-5$ (depending on redshift and
$L_{\rm IR}$) when using 24~\micron\ data alone compared to that
calculated using the full IR SED incorporating longer wavelength data
(at 70~\micron\ and 160~\micron; see \citealt{Papovich07,
    Magnelli09}, Magnelli et al. 2009 [in prep.]).  Furthermore, as
emission from high redshift starbursting systems peaks at longer
wavelengths than those probed by the 24~\micron\ waveband, the sole
use of near and mid-IR observations hampers our ability to efficiently
distinguish between star-formation and AGN dominated systems. This too
can be resolved by incorporating longer wavelength IR data into our
analysis.  Undoubtedly, the launch of the \Herschel\ \textit{Space
  Telescope}, with its ability to observe the universe at
$55-672~\micron$, will help in addressing these issues.  However,
progress can be made now by applying stacking analyses to deep
\Spitzer\ 24~\micron\ and 70~\micron observations, enabling us
to measure the average IR properties of AGNs at a sensitivity level
comparable to that achievable for individual sources in the proposed
deep \Herschel\ fields (e.g. the HGOODS Key Programme\footnote{URL:
  http://herschel.esac.esa.int/Docs/KPOT/GOODS\_Herschel.pdf}; PI:
D. Elbaz).

In this paper we investigate the mid to far-IR properties of X-ray
detected AGNs that lie in the confusion-limited 70~\micron\ and
24~\micron\ {\it Spitzer}-MIPS Far-Infrared Deep Extragalactic Legacy
Survey (FIDEL) and Great Observatories Origins Deep Survey (GOODS). We
identify AGNs using the 1~Ms \textit{Chandra} Deep Field-South X-ray
observations (which have high quality X-ray spectral constraints
  published in \citealt{Tozzi06}) and explore their 70~\micron\
fluxes, luminosities, and 70/24~\micron\ flux ratios (hereafter,
\fratio) as a function of redshift, X-ray luminosity, and X-ray
absorption. As part of our analysis we use the \Swift-BAT sample
  of local, well studied AGNs to develop diagnostics to distinguish
between starburst (hereafter, SB) and AGN dominated systems at $z
\lesssim 1.5$. In the CDF-S field we find evidence that more X-ray
luminous AGNs have warmer, more AGN dominated IR colours and that
\highz, low X-ray luminosity (i.e., $z=1-2$, $L_{\rm X} =
10^{42-43}$~\ergs) AGNs are significantly more IR luminous than their
low redshift counterparts (i.e., $z=0.5-1$, $L_{\rm X} =
10^{42-43}$~\ergs).  This increase in IR luminosity may be interpreted
as tentative evidence of larger dust covering factors in high
redshift AGNs, supporting results from deep X-ray surveys which
  find evidence of larger levels of \Nh\ at similar redshifts
  (e.g. \citealt{LaFranca05,Hasinger08}. However, a proportion of
this increase may also be attributed to higher levels of
star-formation in these systems.  We show that deep, far-IR \Herschel\
surveys will resolve the dominant process behind this increase in IR
luminosity and enable the discrimination between AGN and
star-formation dominated systems out to $z\approx$~5--6.

Throughout this work we adopt $H_{0}=71$~km~s$^{-1}$~Mpc$^{-1}$,
$\Omega_{\rm M}=0.27$, and $\Omega_{\Lambda}=0.73$. We report 1\sig\
errors for detected sources and provide 3\sig\ upper limits for
undetected sources ($<$3\sig).

\begin{figure*}
\includegraphics[angle=0,width=160mm]{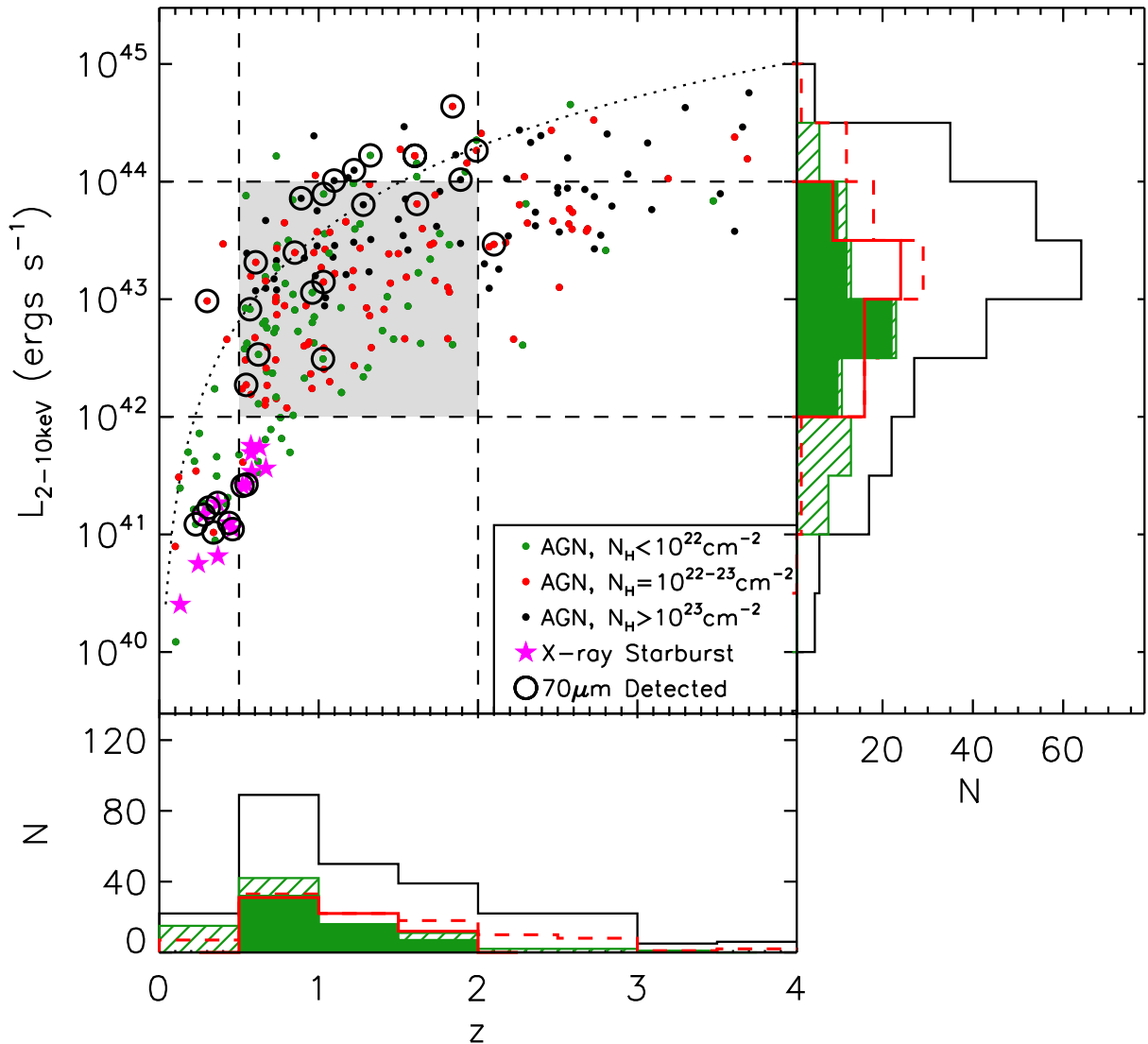}
\caption{\comment{LABEL:LX\_z; } Rest-frame hard X-ray luminosity
  ($L_{\textrm{2-10keV}}$) vs. redshift ({\it z}) of the AGNs and SBs
  identified in the 1~Ms CDF-S X-ray observations, separated according
  to column density (\Nh).  Those sources detected at 70~\micron\ are
  indicated with large circles.  As described in \S\ref{Dat_Xray} this
  sample contains biases in favour of more luminous sources and higher
  column densities at higher redshifts.  To mitigate the effects
  of these biases we define a more heterogeneous subsample (the
  `Restricted' sample), contained within the grey shaded region.
  Coloured histograms indicate the number of AGNs in each redshift and
  X-ray luminosity bin that have column densities $N_{\rm
    H}<10^{22}$~\cs\ (green) and $N_{\rm H}=10^{22-23}$~\cs\
  (red).  The black histogram shows the distribution of the full
  AGN+SB sample.  Solid colours/lines are used to indicate the
  restricted sample, hatching/dashes indicate the (unrestricted, flux
  limited) parent sample.  The curved dotted line indicates the
  constant, unabsorbed hard X-ray flux (6.8\e{-15}~\ergcms) that best
  fits the \Lx-{\it z} distribution of the 70~\micron\ detected X-ray
  AGNs.}
\label{LX_z}
\end{figure*}
 
%
\section{Data}
\label{Data}
%
Our main focus throughout this work is the analysis of the mid and
far-infrared (MIR [5--30~\micron] and FIR [30--300~\micron],
respectively) emission of X-ray detected sources in the CDF-S
field. To aid in the interpretation of the CDF-S data, we also make
use of archival \Spitzer-IRS spectra and \textit{IRAS} data for the
\Swift-BAT sample of local X-ray AGNs, two archetypal AGNs (NGC~1068
and NGC~6240), and a distant quasi-stellar object (QSO) sample. We
describe the CDF-S sample in \S\ref{Dat_CDFS} and the comparison
samples in \S\ref{Dat_Comparison}.

\subsection{CDF-S Sample of X-ray Detected AGNs \& Starbursts}
\label{Dat_CDFS}
\subsubsection{{\it Chandra} Data}
\label{Dat_Xray}
The X-ray data for our main sample are taken from the 1~Ms CDF-S X-ray
observations (\citealt{Giacconi02}), as analysed by
  \cite{Alexander03}.  In total there are 201, 304, and 326 X-ray
detected objects in this region down to limiting 2-8~keV fluxes of
$10^{-15}$, $10^{-16}$ and $10^{-17}$~\ergcms, respectively. The CDF-S
data is deep enough to detect star-formation activity out to $z
\approx 1$. Since the primary aim of this study is to explore the
  IR properties of X-ray detected AGNs, we have used the
  \cite{Bauer04} X-ray source classifications to separate AGNs from
  star-forming galaxies. In the full catalogue of X-ray sources, there
  are 288 X-ray AGNs and 15 X-ray SBs (the remaining 23 are classed as
  either normal galaxies or stars and are excluded from further
  discussion).  The absorption corrected 2--10~keV X-ray luminosities
  (\Lx) and X-ray absorption column densities (\Nh) are taken from
  \cite{Tozzi06}.  Here, we only consider those 266 X-ray AGNs/SBs
  with well defined \Lx\ and \Nh\ measurements.  Spectroscopic and
  photometric redshifts for these 266 X-ray AGNs/SBs were also taken
  from \cite{Tozzi06}; 137 have spectroscopic redshifts (113 of which
  are described as secure, see \citealt{Tozzi06} for details). We use
  photometric redshifts for the remaining 129 X-ray AGNs/SBs that lack
  spectroscopic redshifts.

\begin{figure}
\includegraphics[angle=0,width=85mm]{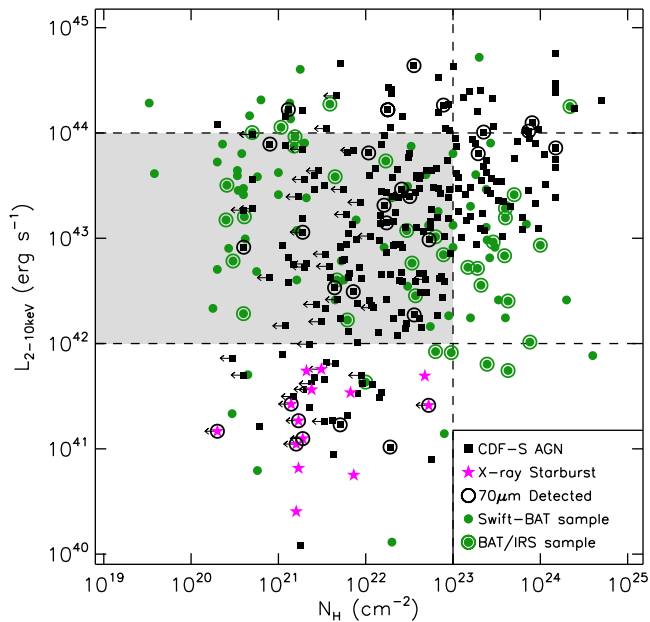}
\caption{\comment{LABEL:LX\_NH; } Rest frame hard X-ray luminosity
  ($L_{\textrm{2-10keV}}$) versus column density (\Nh) of AGNs and SBs
  identified in the 1~Ms CDF-S X-ray observations.  Also shown is the
  \Lx-\Nh\ distribution of the \Swift-BAT sample of $z \approx 0$ AGNs
  which cover the same region of the parameter space as the CDF-S
  sample.  The selection criteria for inclusion in the `Restricted'
  sample is indicated by the shaded region.}
\label{LX_NH}
\end{figure}

In \Fig{LX_z} we present the \Lx-redshift distribution of the 266
X-ray AGNs/SBs considered in this study.  As is to be expected in flux
limited samples, there is a strong bias toward the detection of more
luminous sources at higher redshifts.  Furthermore, as is shown in
\Fig{LX_NH}, there is a bias toward higher column densities at higher
X-ray luminosities because a) only the brightest X-ray sources
can be detected behind $N_{\rm H}>10^{22-23}$~\cs and b) it is
difficult to measure low values of \Nh\ at high redshifts due to the
absorption cut-off being shifted out of the lowest \Chandra\ energy
band.  In our analysis we consider two main samples: a) the full
sample of classified X-ray AGNs/SBs and b) a `Restricted' sample
limited to those AGNs with $L_{\rm X} = 10^{42-44}$~\ergs\ and
$N_{\rm H} < 10^{23}$~\cs\ within the redshift range of $0.5 < z <
2.0$ (indicated by the shaded regions in Figs.\ref{LX_z} and
\ref{LX_NH}). The `Restricted' sample is used to mitigate the effects
of absorption and selection biases on the average results; the $L_{\rm
  X}$ and $N_{\rm H}$ range of the restricted sample is also well
matched to the $z\approx$~0 \Swift-BAT comparison sample (see
\S\ref{Dat_BAT}).

\subsubsection{{\it Spitzer} Data}
The 24~\micron\ and 70~\micron\ {\it Spitzer}-MIPS
GOODS/CDF-S\footnote{URL: http://www.stsci.edu/science/goods/}
observations (PID: 20147; P.I.: D.~Frayer) and FIDEL\footnote{URL:
  http://irsa.ipac.caltech.edu/data/SPITZER/FIDEL/} legacy surveys
(PID: 30948; P.I.: M.~Dickinson) are the deepest available at these
wavelengths and therefore present the best datasets to study the
MIR to FIR emission of AGNs in the \highz\ universe.  While the FIDEL
region covers the full E-CDFS (\citealt{Lehmer05}), the deepest region
is the 10\arcmin~$\times$~10\arcmin\ region centred on the CDF-S field
covered by the GOODS observations. The raw 70~\micron\ data were
processed off-line using the Germanium Reprocessing Tools (GeRT),
following the techniques described in \cite{Frayer06}. The final
mosaics, produced by MOPEX\footnote{URL:
  http://ssc.spitzer.caltech.edu/postbcd/download-mopex.html}, have a
pixel scale of 4.0\arcsec. The 70~\micron\ $3\sigma$ sensitivity is
2.0~mJy in the CDF-S field and $\approx$3.0~mJy in the outer E-CDFS
regions.  The 24~\micron\ images produced using MOPEX have a pixel
scale of 1.2\arcsec. The sky coverage at 24~\micron\ is inhomogeneous
and the sensitivity ranges from 30 to 70~\mujy\ ($5\sigma$). Assuming
a SNR of 4.1 for the the 24~\micron\ observations (corresponding to an
mean flux limit of $\approx$50~\mujy) there are 14025 detected sources
in the FIDEL field.

\subsubsection{X-ray--IR source matching}
\label{Dat_Match}
We matched the classified X-ray sources to the 70~\micron\ sources
assuming a search radius of 2.0\arcsec\ for those sources in the GOODS
field with 24~\micron /IRAC counterparts and 4.0\arcsec\ in the larger
FIDEL field.  This yields 30 matches which are listed in
\Tab{Detected_Info}, representing $\approx$~11\% of the 266 X-ray
AGNs/SBs.  Using the $P$-statistic ($P = 1 - \textnormal{exp}(-\pi n
\theta^2$); e.g.,\ see \citealt{Downes86} and Eqn~1 in
\citealt{Pope06}) and a 70~\micron\ source density
$n\approx$2.0\e{3}~deg$^{-2}$, the chance of one or more 70~\micron\
source lying within 4.0\arcsec of an X-ray source is $P\approx0.2$\%.
Given the small value of $P$, it was not necessary to apply the
correction to the matching procedure described by \cite{Downes86}.

We matched the X-ray sources to the 24~\micron\ sources using a
3.0\arcsec\ search radius; this search radius was found to provide the
best compromise between finding real matches while reducing the number
of spurious matches. Out of the 266 CDF-S X-ray AGN/SBs we found 172
have 24~\micron\ counterparts (i.e.,\ $\approx$~65\%). The number of
spurious matches in the CDF-S field was estimated to be $\sim 4.5\%$
using two approaches (1) by calculating the area of the field covered
by 24~\micron\ sources, assuming that each source is a circle of
radius 3.0\arcsec, and (2) by shifting the positions of the
24~\micron\ sources by a random displacement between 20$\arcsec$ and
55$\arcsec$ and then re-matching to the X-ray sample.  This spurious
matching fraction is consistent with the $P$-statistic ($\approx$3\%),
assuming $n\approx$1.5\e{4}~deg$^{-2}$ (see above). In spite of the
larger $P$-statistic associated with the 24~\micron\ matches compared
to the 70~\micron\ matches, the correction described in
\cite{Downes86} need only be applied to 15 (6\%) of the 266 X-ray
sources and has negligible effect on any of our results.

All but one of the 70~\micron\ detected sources has a 24~\micron\
counterpart. Further investigation reveals that the lack of a matched
24~\micron\ counterpart is the result of the larger PSF of the
70~\micron\ image; at the corresponding position in the 24~\micron\
image there is a cluster of sources that lie just outside the
3.0\arcsec\ matching radius which blend to form a single ``source'' in
the 70~\micron\ image.  To mitigate source blending we de-blend the
70~\micron\ sources with Gaussians placed at the 24~\micron\
positions.  It is these `de-blended' fluxes that are reported in
\Tab{Detected_Info}.

Finally, one of the 70~\micron\ detected X-ray sources lies within
3.0\arcsec\ of two 24~\micron\ sources (\citealt{Tozzi06} Index: 31);
in \Tab{Detected_Info} we list both 24~\micron\ matches but assume
that the brighter of the two 24~\micron\ sources is the real match in
our analysis.

\subsubsection{24~\micron\ and 70~\micron\ stacking procedure}
\label{Dat_Stack}
As the majority of the X-ray sources are not detected at 70~\micron\
we rely on stacking analyses to provide insight into their average MIR
to FIR properties.  Stacking was performed using the code of
\cite{Huynh07}.  Cutouts of 128\arcsec\ x 128\arcsec\ (32 x 32 pixels)
were made from the 70~\micron\ image at the stack positions and
combined using a weighted mean. The 70~\micron\ stacked flux density
was measured using an aperture of 8.0\arcsec\ at the stacked image
center. Offset stacks were generated using random but nearby offset
positions ($< 64$\arcsec) in the 70~\micron\ image. Two hundred
randomly offset stacks were made and the uncertainty in the stacked
flux is taken as the standard deviation of the measured flux
density in the 200 offset stacks.

We created stacks that either include or exclude the 70~\micron\
detected sources; the former (hereafter, \textit{``all''}) provides a
global average of the FIR properties of all X-ray sources and the
latter (hereafter, \textit{``undetected''}) provides the average of
the 70~\micron\ undetected X-ray sources that is less skewed by
individual bright 70~\micron\ sources.  We also stack only the
70~\micron\ detected sources for comparison (hereafter,
\textit{``detected''}).  Because the \textit{all} stacks often appear
to be dominated by the few 70~\micron\ detected sources we generally
use the \textit{detected} and \textit{undetected} stacks in our
analysis.

In order to isolate trends between the IR emission from AGNs and
other physical properties we further split the `Restricted'
sample (see \S\ref{Dat_Xray}) in terms of \Lx\ ($=10^{42-43}$~\ergs,
$=10^{43-44}$~\ergs), \Nh\ ($< 10^{22}$~\cs, $=10^{22-23}$\cs) and
redshift ($z=0.5-1.0$, $z=1.0-2.0$).

\subsection{Comparison Samples}
\label{Dat_Comparison}
In our analysis we compare the results from the sample of CDF-S X-ray
AGNs/SBs to the IR properties of two archetypal AGNs, a sample
of bright QSOs from \cite{Richards06} and, most importantly, a sample
derived from local ($z<0.1$) AGNs in the \Swift-BAT catalogue of hard
X-ray detected AGNs (see \S\ref{Dat_BAT}, appendix and
\citealt{Tueller08}).  For the latter we selected those AGNs that
cover the same range of X-ray properties (i.e. \Lx\ and \Nh; see
\Fig{LX_NH}) as the CDF-S X-ray AGNs sample and also have archival
\Spitzer-IRS spectra and/or \IRAS\ flux density measurements.  The
result is a sample that is \textit{directly comparable} to the more distant CDF-S AGNs, but which has far superior IR data due to
the relative proximity of the sources.

\subsubsection{The \Swift-BAT Sample}
\label{Dat_BAT} 
The \textit{Swift} telescope is currently undertaking a survey of
X-ray bright ($F_{14-195{\rm ~keV}} \gtrsim $ 5\e{-11}~\ergcms) AGNs
using its on-board Burst Alert Telescope (BAT); the second data
release (DR2) of this survey has recently been published in
\cite{Tueller08}.  As the BAT instrument is sensitive only to very
hard X-rays (14-195~keV) this survey has resulted in a sample of
local\footnote{135 of the 153 AGNs with measured redshifts in the BAT
  sample (presented in \citealt{Tueller08}) have $z < 0.1$.} AGNs that
is largely unaffected by absorption (to $N_{\rm H} < 10^{24}$~\cs).
The \Lx\ and \Nh\ distributions of the 104 AGNs for which archival
2-10~keV data exist (\citealt{Bassani99, Winter09}) are comparable to
the CDF-S sample (see Figs. 7 and 10 of \citealt{Winter09}).  Of these
104 AGNs, 61 have flux density measurements from all four \IRAS\ bands
(these 61 objects are hereafter referred to as the BAT/\IRAS\ sample)
which we convert to \Lir\ using Eqns. 2 and 3 in Table 1 of
\cite{Sanders96}.  A search of the \Spitzer\ archives reveals that 36
of the 104 AGNs with 2-10~keV data have \Spitzer-IRS spectroscopy
between 5.2--38~\micron\ (these 36 objects are hereafter referred to
as the BAT/IRS sample) which we use in conjunction with \textit{IRAS}
data to derive the expected observed 24~\micron\ and 70~\micron\
fluxes at the redshifts covered by the CDF-S sample.  All of the AGNs
in the BAT/IRS sample are also in the BAT/IRAS sample. See appendix
for full details of the IRS/\textit{IRAS} analysis.

On separating the BAT/IRS sample into SB and AGN dominated objects we
find that the $S_{70}/S_{24}$ ratio is able to discriminate between
these two types of sources out to $z\approx1.5$ (see shaded regions in
\Fig{Ratios} and appendix).  Furthermore, we note that the 70~\micron\
flux density can predict the infrared luminosities of all the BAT/IRS
objects (i.e., both SB-dominated and AGN-dominated AGNs), as well as
the starburst galaxies described in \cite{Brandl06}, to within a
factor of $\approx$3.  This compares to a factor $\approx$12
uncertainty in \Lir\ using 24~\micron\ flux densities alone (see
appendix and \Fig{W09_Tracks}).

In the plots that follow, we indicate the average IR properties
of the BAT/IRS sample using red and blue lines for the AGN and SB
dominated systems, respectively.  Shading is used to indicate the
range of MIR and FIR properties of the BAT/IRS sample.

\subsubsection{NGC~1068, NGC~6240 and the QSO sample}
In addition to the BAT/IRS sample we also compare the CDF-S AGNs with
two well studied heavily obscured AGNs, NGC~1068 and NGC~6240.  The
former is regarded as the ``quintessential'' type-II AGN, showing
evidence of a hidden broad line region in polarised light
(e.g. \citealt{Antonucci85}); the latter is a heavily obscured
  AGN that is SB-dominated at infrared wavelengths and is often cited to
  characterise the properties of faint X-ray AGNs.  The tracks for
these AGNs were derived from archival \Spitzer-IRS spectra following
the same procedure as used for the BAT/IRS sample as outlined in the
appendix.

Finally, we use the average QSO SED of \cite{Richards06} to give an
indication of the typical IR properties of luminous, unobscured
type-I AGNs.
\\

\noindent
We stress that the comparison samples considered here do not cover the
full range of SEDs seen in the general galaxy population.  It is
known, for example, that very strongly starbursting systems have
larger \fratio\ flux ratios than NGC~6240 (e.g. Arp~220).
Furthermore, some quiescent galaxies are known to have low \fratio\
flux ratios similar to those of AGNs in the BAT/IRS sample (see
Alexander et al., in prep). However, our main aim here is to explore
the IR emission of AGNs with X-ray properties comparable to the
CDF-S AGNs, which excludes such extreme systems.  We find that known
starbursting systems, such as those presented in \cite{Brandl06}, all
produce expected \fratio\ flux ratios either consistent with or higher
than those of the SB-dominated BAT/IRS AGNs, as indicated by the
  shaded region in \Fig{W09_Tracks}.

%
\section{Results}
%

\begin{figure*}
\includegraphics[angle=0,width=160mm]{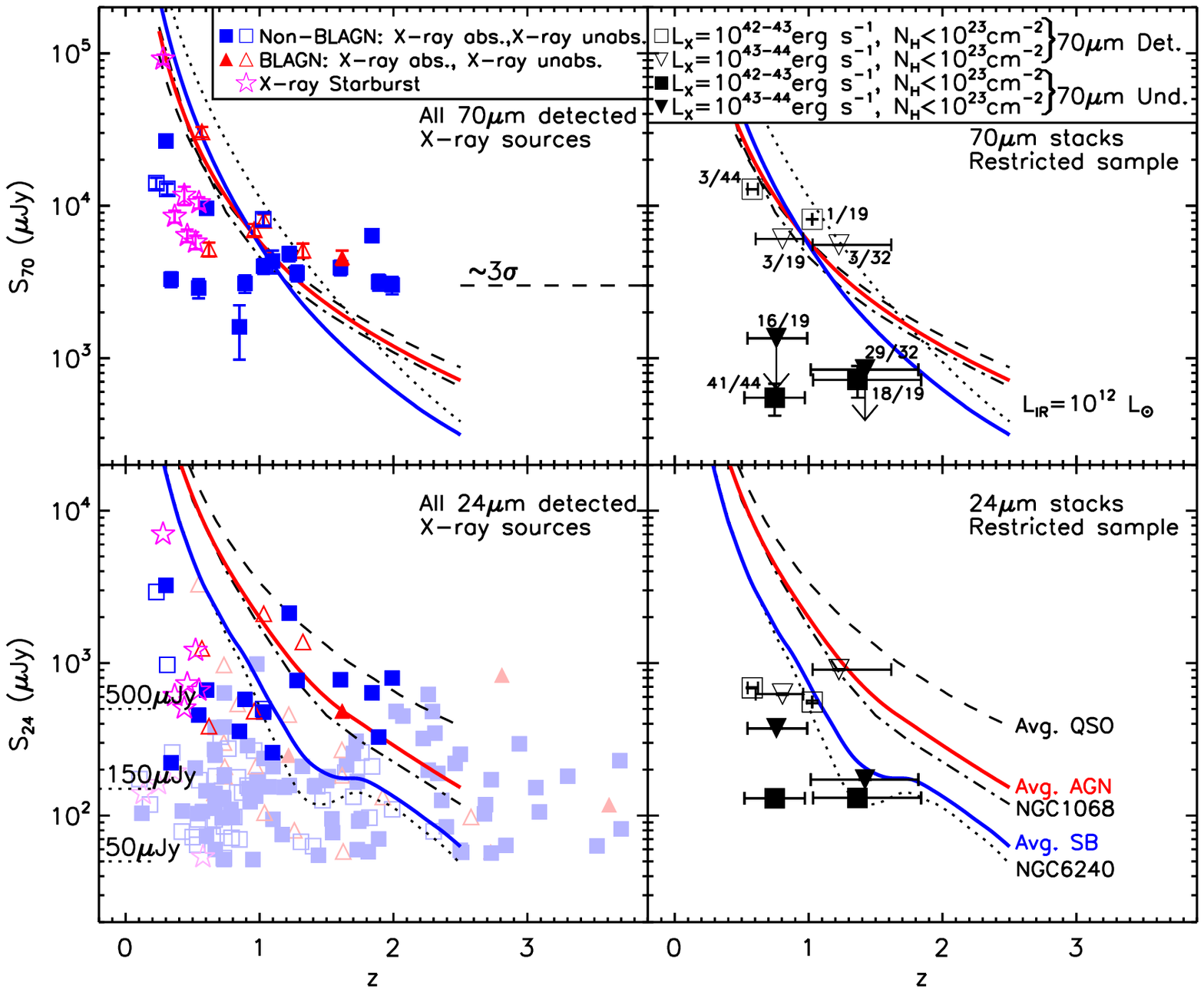}
\caption{\comment{LABEL:Fluxes; }\textit{Left Panels:} 70~\micron\
  (\textit{top}) and 24~\micron\ (\textit{bottom}) flux densities of
  the CDF-S X-ray SBs and AGNs that are detected in both the
  70~\micron\ and 24~\micron\ wavebands (strong colours) and only the
  24~\micron\ waveband (faint colours).  \textit{Right Panels:} The
  average 70~\micron\ ({\it top}) and 24~\micron\ ({\it bottom}) flux
  densities of the 70~\micron\ {\it detected} and 70~\micron\ {\it
    undetected} AGNs in the `Restricted' sample, split according to
  \Lx and {\it z}.  These averages are derived from stacking analyses.
  The fractions indicate the number of AGNs in each stack as well as
  in the whole subsample.  The five tracks in each panel show the
  expected flux densities of the comparison AGNs, normalised to
  \Lir=$10^{12}$~\Lsun, at $z=0.25-2.5$.}
\label{Fluxes}
\end{figure*}

\begin{figure*}
\includegraphics[angle=0,width=160mm]{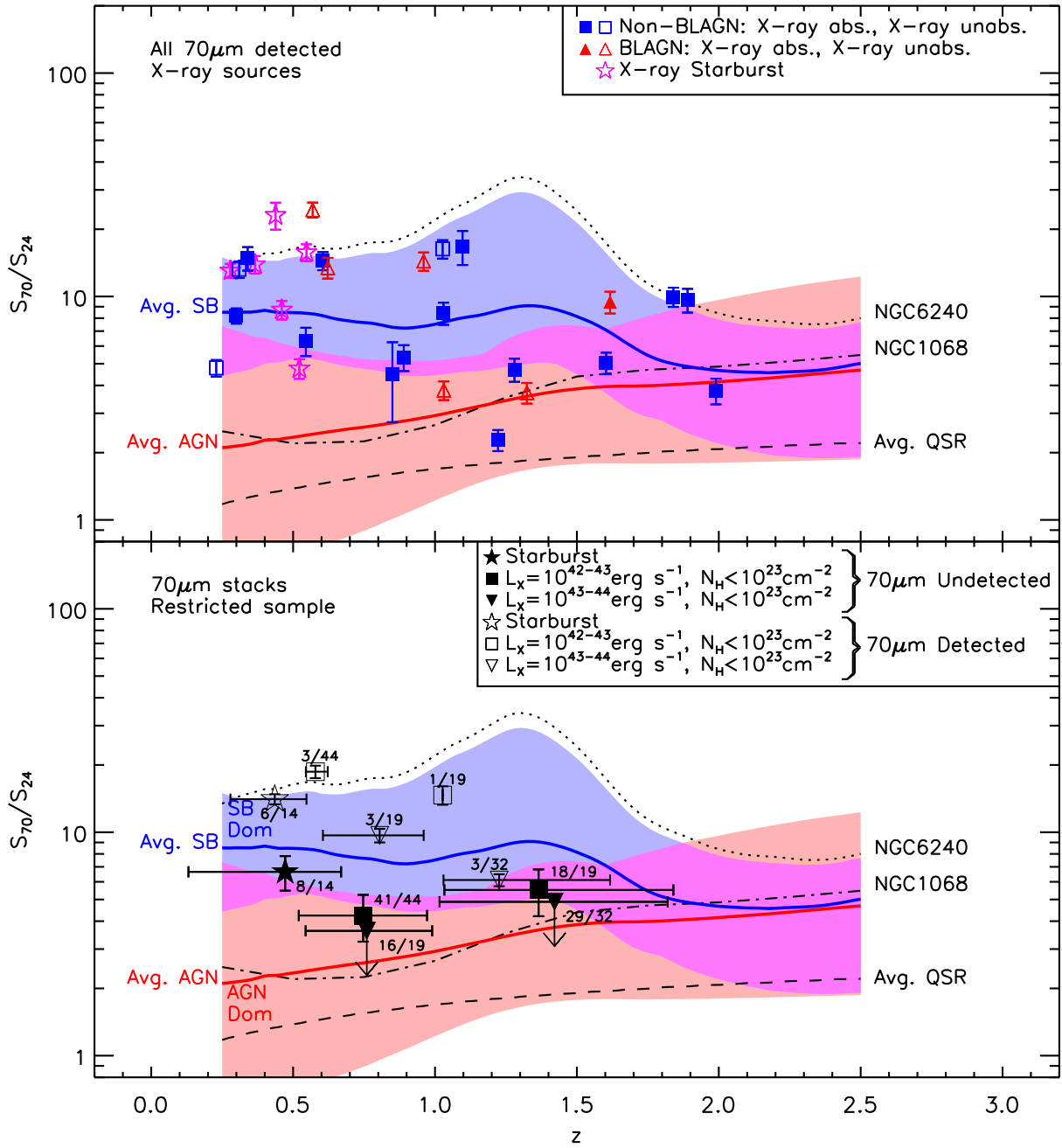}
\caption{\comment{LABEL:Ratios; }\textit{Top:} \fratio\ flux ratios of
  the CDF-S X-ray AGNs/SBs that are detected at 70~\micron.
  \textit{Bottom:} The average \fratio\ flux ratios of the 70~\micron\
  {\it detected} and 70~\micron\ {\it undetected} AGNs in the
  `Restricted' sample, split according to \Lx and {\it z}.  These
  averages are derived from stacking analyses.  The fractions indicate
  the number of AGNs in each stack as well as in the whole subsample.
  The five tracks indicate the expected \fratio\ flux ratios of the
  comparison AGNs at $z=0.25-2.5$; see \S\ref{Dat_Comparison} and
  appendix.  The shaded regions indicate the range of expected
  \fratio\ flux ratios of the AGNs in the BAT/IRS at $z = $ 0.25--2.5.}
\label{Ratios}
\end{figure*}

\begin{figure*}
\includegraphics[angle=0,width=170mm]{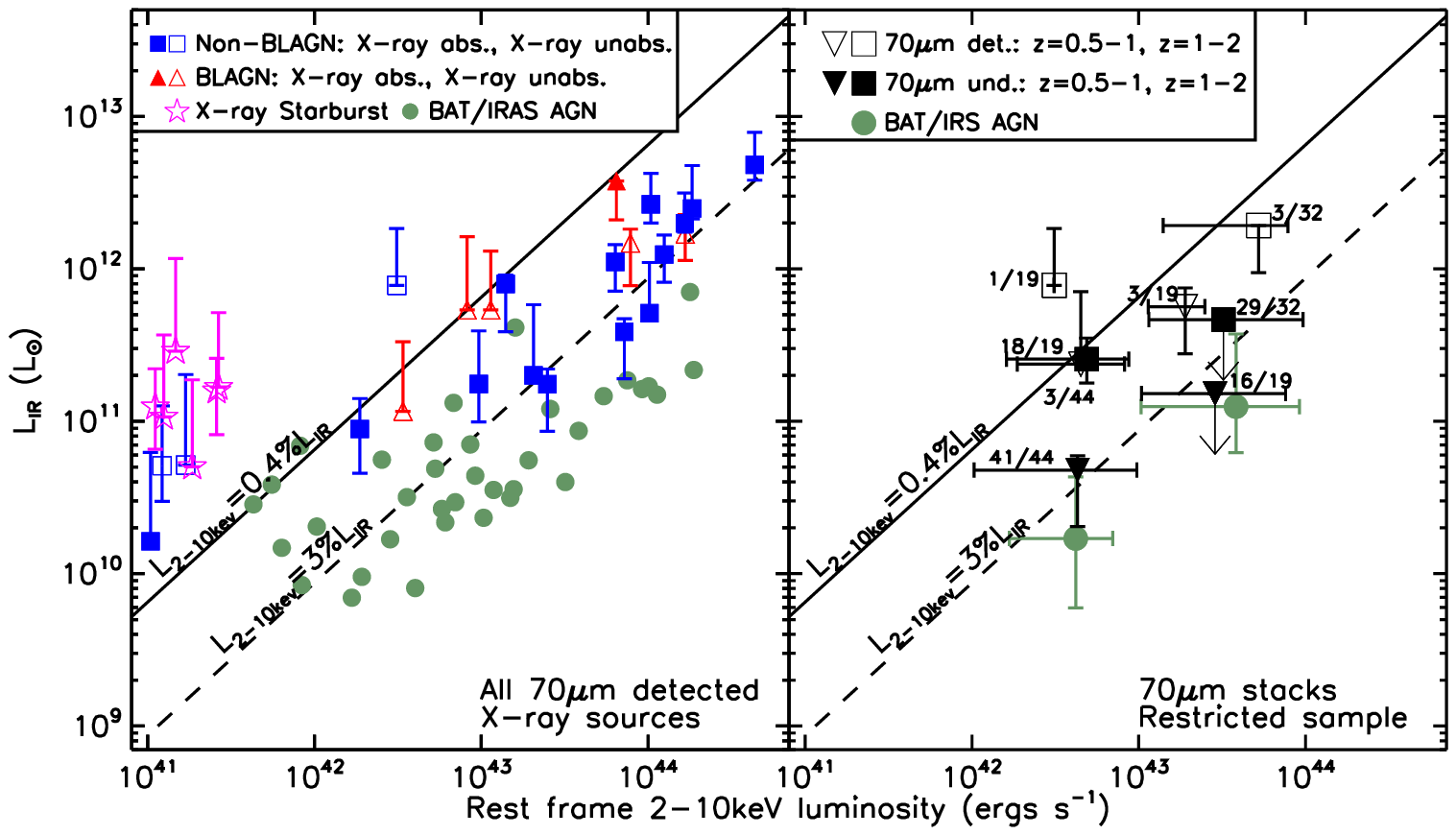}
\caption{\comment{LABEL:LIR\_vs\_LX; } \Lir\ versus \Lx\ of the
  70~\micron\ detected CDF-S X-ray AGNs/SBs ({\it left}) and the
  70~\micron\ {\it undetected} and 70~\micron\ {\it detected} stacks
  of the `Restricted' sample, separated according to redshift
  (\textit{right}).  In both plots, the vertical error bars indicate
  the range of \Lir\ produced by assuming the full range of
  tracks in Figs. \ref{Fluxes} and \ref{Ratios} and horizontal error
  bars indicate the full range of \Lx\ in each stack.  For comparison
  we include the BAT/IRAS AGNs (see \S\ref{Dat_BAT}) and lines
  of constant \Lir/\Lx\ for the average QSO of \protect
  \citeauthor{Elvis94} (\protect \citeyear{Elvis94}; \Lx$=3$\%\Lir)
  and the AGN-hosting submillimeter galaxies of \protect
  \citeauthor{Alexander05} (\protect \citeyear{Alexander05};
  \Lx$=0.4$\%\Lir).  For the average BAT/IRAS points we only include
  BAT/IRAS AGNs with X-ray properties matching those of the
  `Restricted' sample (i.e., \Lx$=10^{42-44}$~\ergs,
  \Nh$<10^{23}$~\cs).}
\label{LIR_vs_LX}
\end{figure*}

\label{Analysis}
In our analysis we investigate the IR properties of the CDF-S X-ray
AGNs and SBs, including the \fratio\ flux ratios, the IR luminosities
and the IR$-$X-ray luminosity ratios.  To aid in our analysis we
compare the properties of the CDF-S X-ray sources with expectations
based on the more detailed analysis of nearby AGNs with the same X-ray
properties as the CDF-S sample (i.e.,\ $L_{\rm X}$; $N_{\rm
  H}$). Since only a minority of CDF-S X-ray AGNs are detected at
70~\micron, our analyses rely significantly on 70~\micron\
stacking analyses and the 24~\micron\ fluxes of individual sources.

\subsection{Individual 70~\micron\ detected sources}
\label{Ana_Match}
\subsubsection{X-Ray properties and redshift distribution}
\label{Ana_Match_Xray}
In \Fig{LX_z} we show the X-ray luminosity-redshift distribution for
the CDF-S X-ray AGNs/SBs and highlight the 70~\micron\ detected
objects.  As might be expected, there is a preference for bright X-ray
AGNs to be detected at 70~\micron.  However, there are up to 2 orders
of magnitude difference in \Lx\ for 70~\micron-detected sources at the
same redshift, suggesting a large range of X-ray/IR luminosity
ratios. As we show in \Fig{Fluxes}, detection at 70~\micron\ is
strongly dependent on the 24~\micron\ flux density.  None of the AGNs
with $S_{24}<200$~\mujy, and fewer than a quarter (10/45;
$\approx$22\%) of those with $200<S_{24}/$\mujy$<500$, are detected at
70~\micron.  Conversely, the majority (19/28; $\approx$68\%) of those
AGNs with $S_{24}>500$~\mujy\ are detected at 70~\micron.

The 70~\micron\ detected sources span a broad range of X-ray
properties. Of the 29 sources detected at both 70~\micron\ and
24~\micron, 15 ($\approx$52\%) are non-broad-line AGNs (non-BLAGNs), 6
($\approx$21\%) are broad-line AGNs (BLAGNs) and 8 ($\approx$28\%) are
X-ray detected SBs. These 70~\micron\ detected sources comprise 15/218
($\approx$7\%) of the non-BLAGNs, 6/34 ($\approx$18\%) of BLAGNs and
6/14 ($\approx$43\%) of the X-ray detected SBs; the bias toward the
BLAGNs and X-ray detected SBs could be due to the BLAGNs and SBs
having high \Lx\ and large IR--X-ray luminosity ratios,
respectively.  The 70~\micron\ detection of AGNs shows no apparent
dependency on absorbing column density: for example, at $z<1.5$, 6/17
($\approx$~35\%) 70~\micron\ detected sources are unabsorbed ($N_{\rm
  H}<10^{22}$~\cs), 5/17 ($\approx$~29\%) are absorbed ($N_{\rm
  H}\approx 10^{22}-10^{23}$~\cs) and 6/17 ($\approx$~35\%) are
heavily absorbed ($N_{\rm H}>10^{23}$~\cs).

\subsubsection{Distinguishing between SB and AGN dominated systems}
\label{Ana_Det_SBAGN}
In \Fig{Ratios} we show the \fratio\ flux ratios of all of the
CDF-S X-ray AGNs/SBs and compare them to the range expected for the
BAT/IRS AGN sample and the well studied local AGNs and quasars; see
section \ref{Dat_BAT} and appendix. The \fratio\ flux ratio provides
an effective discrimination between those AGNs with AGN-dominated or
SB-dominated IR SEDs to $z\approx 1-1.5$.  Using the tracks derived
for the BAT/IRS sample we find that 15/23 ($\approx$65\%) and 5/23
($\approx$22\%) of the 70~\micron\ detected CDF-S AGNs have \fratio\
flux ratios consistent with SB-dominated and AGN-dominated systems,
respectively; the remaining three 70~\micron\ detected AGNs lie in
regions of the \fratio$-z$ plot that are consistent with either SB or
AGN dominated systems. Ten ($\approx$~42\%) of the 70~\micron\
detected CDF-S AGNs have \fratio\ flux ratios within a factor of 1.5
of the NGC~6240 track, five of which are X-ray absorbed and five of
which are X-ray unabsorbed. The maximum \fratio\ flux ratio of the
70~\micron\ detected sources is $24.4\pm1.9$.  This is marginally
inconsistent with \cite{Papovich07} who found 3/30 ($\approx$10\%)
X-ray detected AGNs in their 70~\micron\ selected sample with
\fratio$>30$.  However, the larger area covered by their survey
increases the likelihood of finding rare AGNs with more extreme
\fratio\ flux ratios that are not represented by our comparison
samples; see \S\ref{Dat_Comparison}.

Among the 70~\micron\ detected sources we find no difference in
\fratio\ flux ratio between BLAGNs and non-BLAGNs, nor between X-ray
absorbed and X-ray unabsorbed AGNs (i.e.,\ $N_{\rm H} \geq
10^{22}$~\cs\ and $N_{\rm H} < 10^{22}$~\cs, respectively), implying
that the material that absorbs photons at X-ray and optical
wavelengths is optically thin to rest frame MIR--FIR radiation, see
\S\ref{Ana_Match_Xray}. The average \fratio\ flux ratios and redshifts
of the absorbed (unabsorbed) 70~\micron\ detected AGNs are
$\overline{S_{70}/S_{24}}=8.2\pm4.3$ $(11.8\pm7.3)$ and
$\overline{z}=1.1\pm0.6$ $(0.76\pm0.56)$\footnote{Given errors
    correspond to the standard deviation in the stacked flux},
respectively.

All six 70~\micron\ detected X-ray SBs have \fratio\ flux ratios
consistent with that expected of SB-dominated systems. However, one of
the SBs has a \fratio\ flux ratio larger than that expected from the
BAT/IRS sample and is typical of the extreme SB systems presented in
\cite{Brandl06}, see \Fig{W09_Tracks}.  The 70~\micron\
detected X-ray SBs also span the same range of \fratio\ flux
ratios as the seven 70~\micron\ detected X-ray AGNs that lie in the
same redshift interval (i.e., $z=0.2-0.7$), providing further evidence
that these AGNs have SB-dominated IR SEDs.

\subsubsection{IR fluxes and luminosities of individual, 70~\micron-detected AGNs}
\label{Ana_Det_Lum}
Since we can provide a good characterisation of the basic IR SEDs of
the 70~\micron-detected CDF-S X-ray AGNs, we can accurately estimate
their IR luminosities. We derive $L_{\rm IR}$ for the 70~\micron\
detected sources by first selecting one of the five tracks
plotted in \Fig{Ratios} that best matches the \fratio\ flux ratio of a
given CDF-S X-ray AGN, then scaling its IR luminosity to reproduce
$S_{70}$ of the CDF-S source if observed at the same
redshift. Estimating $L_{\rm IR}$ in this manner assumes that the IR
SEDs of AGNs at the redshifts covered by the CDF-S observations are
similar to those seen in the comparison AGN samples. We take some
confidence that this is indeed the case as the range of \fratio\ flux
ratios of the 70~\micron\ detected subsample is largely bounded by the
tracks of our comparison sample.

In the left panel of \Fig{LIR_vs_LX} we show the $L_{\rm IR}$ of the
70~\micron\ detected sample plotted as a function of $L_{\rm X}$.  We
find no significant difference in \Lir\ between BLAGNs and non-BLAGNs
or between X-ray absorbed and unabsorbed AGNs. This independence of
\Lir\ on classification provides further supporting evidence that the
material that absorbs photons at X-ray and optical wavelengths is
optically thin to rest frame MIR--FIR radiation, see
\S\S\ref{Ana_Match_Xray} \& \ref{Ana_Det_SBAGN}.  The 70~\micron\
detected CDF-S X-ray AGNs are typically more IR luminous than their
counterparts in the BAT/\IRAS\ sample of local AGN. This result may be
due to the low sensitivity of the 70~\micron\ observations, which
could lead to a bias towards the detection of the most IR-luminous
AGNs; indeed, the range of IR luminosities ($\approx$~2 orders of
magnitude) for the CDF-S X-ray AGNs is narrower than the range of
X-ray luminosities ($\approx$~3 orders of magnitude).  Overall, the
$L_{\rm IR}/L_{\rm X}$ ratios of the 70~\micron\ detected CDF-S
AGNs/SBs span almost 2 orders of magnitude.

The 70~\micron\ detected X-ray SBs have higher \lratio\ ratios than
the majority (20/22; $\approx$ 91\%) of the 70~\micron\ detected X-ray
AGNs.  We find all three 70~\micron\ detected X-ray AGNs with $L_{\rm
  X}<10^{41}$~\ergs\ have \lratio\ ratios and \fratio\ flux ratios
comparable to the 70~\micron\ detected X-ray SBs (\citealt{Tozzi06}
index: 525, 538, 575; see \Tab{Detected_Info}).  However, two of these
sources (index: 525, 538) have optical spectra consistent with AGNs
(\citealt{Szokoly04}) while the remainder (index: 575) has a flat
X-ray spectral index in the deeper 2~Ms CDF-S catalogue ($\Gamma<0.1$;
\citealt{Luo08}).  While this confirms their classification as AGNs,
it is clear that their IR SEDs are SB dominated.

As only a small fraction of the CDF-S X-ray AGN sample is detected at
70~\micron\ we are unable to reliably constrain the IR
properties of the majority of the CDF-S X-ray AGNs/SBs using the
70~\micron\ data alone. To better characterise the IR properties
for the majority of the CDF-S X-ray AGNs we therefore use stacking
analyses.

\subsection{24~\micron\ and 70~\micron\ stacking analysis}
\label{Ana_Stack}
The advantage of stacking analyses is that it provides the average
properties of sources that lie below the individual source detection
limit. However, naturally, information is lost on the fluxes of
individual sources. We also stacked the 70~\micron\ and 24~\micron\
data following the procedure outlined in \S\ref{Dat_Stack} and present
the results in \Tab{Stack_Info}; although the majority of the CDF-S
X-ray sources are detected at 24~\micron, we stacked the 24~\micron\
data to provide flux density constraints consistent with those
obtained at 70~\micron.  For completeness, we report the results from
stacking the whole sample as well as a range of subsamples (split into
bins of \Lx, \Nh, $z$, 24~\micron\ flux density, and object
classification). As noted in \S\ref{Dat_Stack}, we stacked all X-ray
sources in each subsample ({\it ``all''}), and also stacked only the
70~\micron\ detected objects ({\it ``detected''}) and only the
70~\micron\ undetected objects ({\it ``undetected''}). The majority of
our analyses will focus on the `Restricted' sample to reduce selection
and sensitivity biases (see \S\ref{Dat_Xray}).


All of the CDF-S X-ray sample is detected significantly in the
70~\micron\ stacks, with an average flux density of
$S_{70}$=4824~$\pm$~210~\mujy. Significant detections ($S/N>3$) are
also found from 70~\micron\ stacking analyses for the majority of the
subsamples; however, we note that only 9 ($\approx$~50\%) of the
eighteen 70~\micron\ {\it undetected} subsamples are significantly
detected. From stacking the CDF-S X-ray sample in bins of 24~\micron\
flux density (i.e. $S_{24}$=50-150~\mujy, 150-500~\mujy\ and
$\geq$500~\mujy), we find that the 70~\micron\ flux density and the
significance of the detection is positively correlated with the
24~\micron\ flux density of the stacked sources, confirming our
previous result for the 70~\micron\ detected sources; see
\S\ref{Ana_Match}.

We find that when the `Restricted' sample is stacked in bins of \Nh\
and $z$ (i.e.,\ stacks 5, 6, 7 and 8 in \Tab{Stack_Info}) only one
70~\micron\ \textit{undetected} stack is detected at 70~\micron\ at a
significance $>3\sigma$. The upper limits for these non-detections
reveal no significant differences in the IR properties of these
subsamples; they are all consistent with \fratio$\approx$4 and
\Lir$\approx10^{11}$~\Lsun.  The lack of any significant difference
between these stacks may be interpreted as providing further tentative
evidence that the material that absorbs X-rays is optically thin to
rest frame MIR--FIR radiation, see \S\S\ref{Ana_Match_Xray},
\ref{Ana_Det_SBAGN} \& \ref{Ana_Det_Lum}. Similarly, none of the
$L_{\rm X}=10^{43-44}$~\ergs, 70~\micron\ \textit{undetected} stacks
(rows 13 and 14 of \Tab{Stack_Info}) are detected at $>3\sigma$ in
either redshift bin ($z=$~0.5--1 and $z=$~1--2). However, in this case
the upper limits \textit{do} provide significant insight into the IR
properties of the CDF-S X-ray AGNs, revealing that high \Lx\ AGNs
have, on average, warmer \fratio\ flux ratios than their low \Lx\
counterparts (as discussed in \S\ref{Ana_Stack_Ratio}).

We proceed with the analysis of the stacked data following a similar
procedure as outlined in \S\ref{Ana_Match} for the 70~\micron\
detected sources.


\subsubsection{SB/AGN contribution to the average IR emission of
  AGNs}
\label{Ana_Stack_Ratio}
In the lower panel of \Fig{Ratios} we show the \fratio\ flux ratios of
the `Restricted' subsamples split in terms of \Lx\ and $z$ (rows 11,
12, 13 and 14 of \Tab{Stack_Info}).  We find that the average \fratio\
flux ratios of all four 70~\micron\ \textit{undetected} stacks lie
within the range of \fratio\ flux ratios expected for the BAT/IRS AGNs
out to $z\approx$~2.5, and run roughly parallel to the average
AGN-dominated and average SB-dominated tracks (see \S\ref{Dat_BAT} and
appendix). The simplest interpretation of this result is that there is
little change in the average IR colour with redshift for AGNs in both
the $L_{\rm X}=10^{42-43}$~\ergs\ and $L_{\rm X}=10^{43-44}$~\ergs\
bins. However, the conservative upper limits on the \fratio\ flux
ratios of the $L_{\rm X}=10^{43-44}$~\ergs\ 70~\micron\
\textit{undetected} AGNs place them below those of the $L_{\rm
  X}=10^{42-43}$~\ergs\ 70~\micron\ \textit{undetected} AGNs in each
redshift bin. Based on the BAT/IRS tracks this result suggests that
the more X-ray luminous AGNs may have, on average, more
AGN-dominated IR SEDs than their lower \Lx\ counterparts;
  however, deeper data will be required to confirm this
  result. However, we find that the stacks of the $L_{\rm
  X}=10^{42-43}$~\ergs\ AGNs lie in ambiguous regions of the
\fratio$-z$ plot, which limits the conclusions that can be directly
derived on the relative contributions from AGN and SB activity to
these stacks using the \fratio\ flux ratios alone; in
\S\ref{Dis_SB_AGN} we explore a variety of approaches to
constrain the relative AGN and SB contributions using additional data.



\subsubsection{Average IR fluxes and luminosities}
\label{Ana_Stack_Lum}
The average IR luminosities of the stacked subsamples are calculated
by taking the same approach as that used for the individual
70~\micron\ detected sources (see \S\ref{Ana_Det_Lum}).  For those
stacks with less than 3\sig\ detections we use the nominal \fratio\
flux ratio (rather than upper limits) when determining the closest
match out of the five tracks considered; although we note that using
upper limits to select the appropriate track changes the estimates of
\Lir\ by less than 20\%. In both redshift bins the nominal \fratio\
flux ratios of the \Lx$=10^{42-43}$~\ergs\ and \Lx$=10^{43-44}$~\ergs\
70~\micron\ \textit{undetected} stacks are most closely matched by the
average AGN and average QSO SEDs, respectively.  We note, however,
that the 3~\sig\ upper limits on the \fratio\ flux ratios of stacked
\Lx$=10^{43-44}$~\ergs\ AGNs are also consistent with the average AGN
SED; see \Fig{Ratios}.

\begin{figure}
\includegraphics[angle=0,width=85mm]{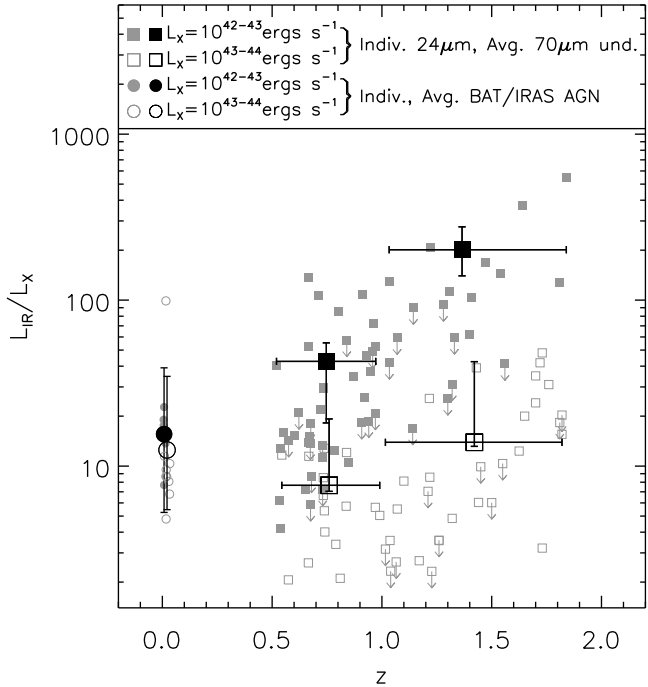}
\caption{\comment{LABEL:LIRLX\_vs\_z;} $L_{\rm IR}/L_{X}$ versus
  redshift for the stacked 70~\micron\ \textit{undetected} sources in
  the `Restricted' sample of CDF-S X-ray AGNs, separated according to
  \Lx\ (black squares).  Vertical error bars indicate the range of
  \lratio\ produced by assuming the various tracks in
  Figs. \ref{Fluxes} and \ref{Ratios}.  Small, grey squares indicate
  the \lratio\ ratios of each X-ray AGN in the `Restricted' sample
  derived from their 24~\micron\ flux densities, assuming the
    Average AGN and Average QSO SEDs for the $L_{\rm
      X}=10^{42-43}$~\ergs\ and $L_{\rm X}=10^{43-44}$~\ergs\ AGNs,
    respectively (see \S\ref{Dis_24um}).  For those sources
  undetected at 24~\micron\ we derive upper limits by assuming a
  24~\micron\ flux density limit of 50~\mujy.  For comparison we
  include the individual and average \lratio\ ratios of the BAT/IRAS
  sample, separated according to \Lx\ (vertical error bars indicate
  the standard deviation of \lratio\ of the BAT/IRS sample).
    Both the stacking analysis and the individual 24~\micron\
    detections clearly show a consistent increase in the \lratio\
    ratio of $L_{\rm X}=10^{42-43}$~\ergs\ AGN from $z\approx0$ to
    $z=1-2$.}
\label{LIRLX_vs_z}
\end{figure}

In \Fig{LIR_vs_LX} we plot the IR luminosity versus \Lx\ for the CDF-S
X-ray sources and the BAT/\IRAS\ sample. In the case of the $L_{\rm
  X}=10^{42-43}$~\ergs\ \textit{undetected} AGNs, we find that $L_{\rm
  IR}$ is larger by a factor of $\approx 5$ for the $z=1.0-2.0$ AGNs
when compared to the $z=0.5-1.0$ AGNs ($L_{\rm IR}=$2.5\e{11}~\Lsun\
and 5.0\e{10}~\Lsun, respectively), and a factor of $\approx$20 times
larger than the average $42<$log(\Lx/\ergs)$<43$, \Nh$<10^{23}$~\cs\
BAT/IRS AGN (i.e., at $z\approx 0$; $L_{\rm IR}=$1.3\e{10}~\Lsun).
This is in spite of these subsamples having almost the same average
X-ray luminosities ($\overline{L_{\rm X}} = $4.9\e{42},
4.3\e{42}~\ergs and 3.7\e{42}~\ergs for $z=1.0-2.0$, $z=0.5-1.0$ and
$z\approx0$ AGNs, respectively). This difference corresponds to a
factor of $4.7_{-2.0}^{+10.2}$ and $12.7^{+7.1}_{-2.6}$ increase in
\lratio\ from $z=0.5-1$ to $z=1-2$ and from $z\approx0$ to
$z=1.0-2.0$, respectively (these conservative errors correspond to the
full range of IR luminosities derived using all five tracks in
\Fig{Ratios}).  The stacked data therefore suggest that the average
\lratio\ ratio of $L_{\rm X}=10^{42-43}$~\ergs\ AGNs strongly evolves
with redshift, as illustrated in \Fig{LIRLX_vs_z}. Although this
result is based on stacking analyses, we show in \S\ref{Dis_24um} that
we obtain the same result using 24~\micron\ constraints for individual
sources.

The 70~\micron\ {\it undetected} stacks of the $L_{\rm
  X}=10^{43-44}$~\ergs\ AGNs at both $z=$~0.5--1 and $z=$~1--2 are not
significantly detected, limiting the constraints that we can place on
$L_{\rm IR}$ using the 70~\micron\ data alone; the limits are $L_{\rm
  IR}<$1.5\e{11}~\Lsun\ and $<$4.6\e{11}~\Lsun\ for the AGNs at
$z=$~0.5--1 and $z=$~1--2, respectively (these upper limits include
the uncertainty in the $S_{70}$--\Lir\ correction).  However, because
the upper limits on the \fratio\ flux ratios of these stacks provide
significant constraints to the range of potential matching BAT/IRS AGN
SEDs we can estimate the average IR luminosities from the stacked
24~\micron\ flux densities to a much higher degree of accuracy than
would normally be the case (to within a factor of $\approx$3, rather
than $\approx$12). This reduced uncertainty is based on the fact that
the \fratio\ flux ratios of the $L_{\rm X}=10^{43-44}$~\ergs\ AGN
stacks restricts us to AGN-dominated BAT/IRS SEDs, which all have
similar $S_{24}$--$L_{\rm IR}$ ratios.  Using the stacked 24~\micron\
flux density we obtain \Lir=5.6\e{10}~\Lsun\ and 1.2\e{11}~\Lsun\ for
the $L_{\rm X}=10^{43-44}$~\ergs\ AGNs in the $z=0.5-1.0$ and
$z=1.0-2.0$ redshift bins, respectively, corresponding to a factor of
$\approx$2 increase between these redshift ranges.  For comparison,
the average \Lir\ of $L_{\rm X}=10^{43-44}$~\ergs\ AGNs in the $z
\approx 0$ BAT/IRS sample is 1.2\e{11}~\Lsun. Given the average X-ray
luminosities of these AGNs (i.e., $\overline{L_{\rm X}} =
$3.8\e{43}~\ergs, 2.9\e{43}~\ergs\ and 3.2\e{43}~\ergs\ for the
$z\approx$~0, $z=$~0.5--1 and $z=$~1--2 AGNs, respectively), we
constrain the change in \lratio\ to a factor of $\approx
1.8^{+4.2}_{-1.0}$ between $z=$~0.5--1 and $z=$~1--2 and to a factor
of $\approx 1.2^{+2.6}_{-0.8}$ between $z \approx 0$ and $z=$~1--2,
both of which are consistent with no evolution; see \Fig{LIRLX_vs_z}.

A number of recent studies have shown that the MIR emission of AGNs
can be used as a good proxy for their intrinsic luminosities
(e.g. \citealt{Krabbe01, Lutz04, Maiolino07, Horst08}).  Understanding
how the X-ray--MIR relationship changes with redshift would be very
useful for detecting potentially heavily obscured AGNs (e.g.,
\citealt{Daddi07, Fiore08, Georgantopoulos08}).  However, previous
X-ray--MIR studies of AGNs have been limited to extremely high
luminosity sources (i.e., $\overline{L_{\rm X}}=$3\e{45}~\ergs) at
redshifts beyond $z \approx 0.3$.  Using the results from our stacking
analyses we can constrain the X-ray--MIR luminosity ratios (hereafter
we focus on \vlvratio) of more typical, distant AGNs found in deep,
multi-wavelength surveys.  At $z\approx 0$ the \lsixratio\ ratios of
the Average AGN and Average QSO SEDS are consistent with \cite{Lutz04}
(i.e., $\approx 3$ and $\approx 2$, respectively)\footnote{We also
  note that the $\nu L_{\nu}(10.5\mu m)/L_{\rm X}$ ratios of the Average
  AGN and Average QSO at $z\approx 0$ are largely consistent with the
  relationship presented in \cite{Krabbe01} (i.e., $\nu
  L_{\nu}(10.5\mu m)/L_{\rm X}\approx 2.5$ and $\approx 1.9$,
  respectively).}.  To calculate the \vlvratio\ ratios of the CDF-S
AGNs we assume the same SEDs as those used to derive their \lratio\
ratios (i.e., Average AGN and Average QSO for $L_{\rm
  X}=10^{42-43}$~\ergs\ and $L_{\rm X}=10^{43-44}$~\ergs\ AGNs,
respectively).  We therefore find that the average \vlvratio\ ratios
of $L_{\rm X}=10^{42-43}$~\ergs\ AGNs increase from $\approx 3$ at $z
\approx 0$ to $\approx 10$ and $\approx 43$ at $z=0.5-1$ and $z=1-2$,
respectively.  By contrast, the average \vlvratio\ ratios of $L_{\rm
  X}=10^{43-44}$~\ergs\ AGNs remain largely unchanged to $z \approx 2$
(i.e., \vlvratio$\approx$~1--2).  Our data therefore suggest that deep
IR observations can be used as a reliable method of identifying
intrinsically bright (i.e., \Lx$=10^{43-44}$~\ergs), potentially
Compton-thick AGNs.  However, we urge caution when using either the
\lratio\ or \vlvratio\ ratios to locate less luminous obscured AGNs
(i.e., $L_{\rm X}\lesssim 10^{43}$~\ergs) unless other indicators of
luminous AGN activity are present (e.g., emission line spectra;
see \citealt{Alexander08}).
 
\clearpage
\pagestyle{empty}
\begin{landscape}
\begin{table}
  \begin{center}{
      \caption{\comment{LABEL:Detected\_Info;} The 30 X-ray detected
        sources with $>3\sigma$ detection at 70~\micron.}
      \begin{tabular}{@{}cclccccccccccl@{}}
\hline
\hline
(1)&(2)&(3)&(4)&(5)&(6)&(7)&(8)&(9)&(10)&(11)&(12)&(13)&(14)\\
T06&A03&Name&\textit{RA}&\textit{Dec}&$z$&$z$ flag&log($L_X$) (ergs s$^{-1})$&$N_H$ ($10^{22}\ \textnormal{cm}^{-2})$&$S_{70}$ $(\mu\textnormal{Jy})$&$S_{24}$ $(\mu\textnormal{Jy})$&$\frac{S_{70}}{S_{24}}$&$L_{IR}$ ($L_\odot$)&Classification\\
\hline
       7&     312&J033301.8-275818& 53.2574&-27.9718&1.840&0.6&44.64&3.57&6370 $\pm$ 620&640 $\pm$ 11&9.96 $\pm$ 0.99&12.7&Abs. Non-BLAGN \\
      29&     240&J033239.0-275701& 53.1626&-27.9505&0.300&0.9&42.98&5.32&26600 $\pm$ 2100&3239 $\pm$ 11&8.20 $\pm$ 0.64&11.2&Abs. Non-BLAGN \\
      31&     230&J033237.9-275213& 53.1577&-27.8704&1.603&3.0&44.22&1.79&3920 $\pm$ 430&777.4 $\pm$ 9.3&5.05 $\pm$ 0.56&12.3&Abs. Non-BLAGN \\
      31&     230&J033237.9-275213& 53.1577&-27.8704&1.603&3.0&44.22&1.79&3920 $\pm$ 430&248.3 $\pm$ 9.3&15.8 $\pm$ 1.8&12.2&Abs. Non-BLAGN \\
      36&     204&J033233.1-274548& 53.1379&-27.7634&1.030&0.5&43.15&1.73&4020 $\pm$ 440&477 $\pm$ 13&8.42 $\pm$ 0.96&11.9&Abs. Non-BLAGN \\
      44&     173&J033226.6-274036& 53.1108&-27.6767&1.031&3.0&43.89&0.08&8010 $\pm$ 770&2107 $\pm$ 15&3.80 $\pm$ 0.37&12.2&Unabs. BLAGN     \\
      46&     163&J033225.3-274219& 53.1052&-27.7054&1.617&3.0&43.81&1.08&4580 $\pm$ 500&485 $\pm$ 13&9.4 $\pm$ 1.1&12.6&Abs. BLAGN     \\
      51&     118&J033217.3-275221& 53.0719&-27.8728&1.097&3.0&44.01&22.42&4340 $\pm$ 740&259.2 $\pm$ 8.6&16.7 $\pm$ 2.9&11.7&Abs. Non-BLAGN \\
      52&     117&J033217.2-274304& 53.0718&-27.7178&0.569&3.0&42.91&0.04&30500 $\pm$ 2300&1247 $\pm$ 10&24.4 $\pm$ 1.9&11.7&Unabs. BLAGN     \\
      56&      88&J033213.3-274241& 53.0554&-27.7115&0.605&3.0&43.31&1.62&9620 $\pm$ 860&665 $\pm$ 16&14.5 $\pm$ 1.3&11.3&Abs. Non-BLAGN \\
      69&      38&J033201.5-274138& 53.0063&-27.6941&0.850&0.4&43.40&3.20&1600 $\pm$ 620&356.9 $\pm$ 8.9&4.5 $\pm$ 1.8&11.2&Abs. Non-BLAGN \\
      72&      30&J033158.3-275043& 52.9931&-27.8453&1.990&0.5&44.26&7.77&3020 $\pm$ 400&796.2 $\pm$ 9.1&3.79 $\pm$ 0.50&12.4&Abs. Non-BLAGN \\
      77&     311&J033301.7-274543& 53.2570&-27.7620&0.622&3.0&42.53&0.44&5190 $\pm$ 540&385.7 $\pm$ 9.1&13.5 $\pm$ 1.4&11.1&Unabs. BLAGN     \\
      78&     193&J033230.1-274524& 53.1256&-27.7568&0.960&3.0&43.06&$<$0.19&6960 $\pm$ 640&484 $\pm$ 12&14.4 $\pm$ 1.4&11.7&Unabs. BLAGN     \\
      98&     260&J033244.3-275142& 53.1846&-27.8618&0.279&3.0&41.17&$<$0.02&91600 $\pm$ 6700&7020 $\pm$ 11&13.05 $\pm$ 0.95&11.5&Unabs. Starburst \\
     122&     295&J033257.7-274549& 53.2404&-27.7636&2.100&0.5&43.47&2.58&4030 $\pm$ 460&-&-&-&Abs. Non-BLAGN \\
     152&     302&J033259.4-274859& 53.2476&-27.8165&1.280&0.6&43.80&19.41&3620 $\pm$ 430&769 $\pm$ 11&4.71 $\pm$ 0.56&12.0&Abs. Non-BLAGN \\
     155&      60&J033208.0-274240& 53.0335&-27.7111&0.545&3.0&42.27&3.59&2880 $\pm$ 410&455 $\pm$ 10&6.33 $\pm$ 0.92&10.9&Abs. Non-BLAGN \\
     175&     284&J033252.0-274437& 53.2165&-27.7438&0.522&3.0&41.41&$<$5.26&5770 $\pm$ 580&1213 $\pm$ 11&4.76 $\pm$ 0.48&11.2&Abs. Starburst \\
     206&     109&J033216.3-273930& 53.0678&-27.6586&1.324&3.0&44.22&0.13&5100 $\pm$ 540&1376 $\pm$ 12&3.71 $\pm$ 0.39&12.2&Unabs. BLAGN     \\
     242&     285&J033251.9-274229& 53.2164&-27.7083&1.027&3.0&42.49&0.72&8150 $\pm$ 750&500 $\pm$ 11&16.3 $\pm$ 1.5&11.9&Unabs. Non-BLAGN \\
     253&     131&J033220.1-274448& 53.0839&-27.7467&1.890&1.9&44.02&73.51&3160 $\pm$ 380&327.20 $\pm$ 0.20&9.6 $\pm$ 1.2&12.4&Abs. Non-BLAGN \\
     268&     278&J033249.3-274050& 53.2056&-27.6806&1.222&3.0&44.10&80.44&4850 $\pm$ 520&2127 $\pm$ 16&2.28 $\pm$ 0.25&12.1&Abs. Non-BLAGN \\
     525&     129&J033219.9-274123& 53.0828&-27.6899&0.229&3.0&41.09&0.00&14000 $\pm$ 1200&2932 $\pm$ 17&4.79 $\pm$ 0.41&10.7&Unabs. Non-BLAGN \\
     538&      65&J033208.6-274649& 53.0359&-27.7803&0.310&3.0&41.23&0.51&12900 $\pm$ 1100&973 $\pm$ 14&13.2 $\pm$ 1.1&10.7&Unabs. Non-BLAGN \\
     553&     292&J033256.8-275318& 53.2366&-27.8885&0.366&3.0&41.27&$<$0.17&8520 $\pm$ 770&612.3 $\pm$ 7.1&13.9 $\pm$ 1.3&10.7&Unabs. Starburst \\
     567&     236&J033238.9-274732& 53.1620&-27.7925&0.460&3.0&41.05&$<$0.16&6370 $\pm$ 600&730.9 $\pm$ 8.9&8.72 $\pm$ 0.83&11.1&Unabs. Starburst \\
     575&     115&J033217.2-274922& 53.0715&-27.8230&0.340&3.0&41.02&1.90&3290 $\pm$ 370&221.5 $\pm$ 9.5&14.8 $\pm$ 1.8&10.2&Abs. Non-BLAGN \\
     577&     224&J033236.3-274933& 53.1512&-27.8258&0.547&3.0&41.42&$<$0.14&10470 $\pm$ 890&664 $\pm$ 10&15.8 $\pm$ 1.4&11.2&Unabs. Starburst \\
     608&     318&J033304.0-275027& 53.2665&-27.8409&0.890&3.0&43.86&150.00&3090 $\pm$ 410&578.0 $\pm$ 9.3&5.34 $\pm$ 0.71&11.6&Abs. Non-BLAGN \\
     646&     266&J033245.2-274724& 53.1884&-27.7903&0.438&3.0&41.10&$<$0.19&11700 $\pm$ 1600&507.2 $\pm$ 8.6&23.1 $\pm$ 3.2&11.0&Unabs. Starburst \\
\hline
\end{tabular}

\label{Detected_Info}   
}\end{center} {\sc Notes}: (1) Source index from \cite{Tozzi06}. (2)
Source index from \cite{Alexander03}. (3) Source name. (4-5) X-ray
position (J2000). (6) Source redshift. (7) Redshift flag. (8) Intrinsic
(i.e., corrected for absorption) $2-10 keV$ luminosity. (9) Column
density in units of $10^{22} \textnormal{cm}^{-2}$. (10) 70 \micron\
flux. (11) 24 \micron\ flux. (14) 70/24 \micron\ flux ratio. (13)
$L_{\rm IR}$ derived from the 70 \micron\ flux using the method
described in the text. (14) Classification from \cite{Bauer04}.
Information in cols. (1),(3)-(9) is taken from \cite{Tozzi06}.
\end{table}
\end{landscape}

\clearpage
\pagestyle{empty}
\begin{landscape}
\begin{table}
\begin{center}{
\caption{\comment{LABEL:Stack\_Info;}Descriptions and average properties of the
AGN and SB stacks.}
\begin{tabular}{@{}clcccccccccccccccccc@{}}
\hline
\hline
&&\multicolumn{3}{c}{(3)}&\multicolumn{3}{c}{(4)}&\multicolumn{3}{c}{(5)}&\multicolumn{3}{c}{(6)}&\multicolumn{3}{c}{(7)}&\multicolumn{3}{c}{(8)}\\
(1)&(2)&\multicolumn{3}{c}{$N$}&\multicolumn{3}{c}{Mean z}&\multicolumn{3}{c}{log($\overline{L_X}$) (ergs s$^{-1}$)}&\multicolumn{3}{c}{Stacked $S_{70}$ ($\mu\textnormal{Jy}$)}&\multicolumn{3}{c}{Stacked $S_{24}$ ($\mu\textnormal{Jy}$)}&\multicolumn{3}{c}{$L_{IR}$ $(L_\odot)$}\\
Index&Description&All&Det.&Und.&All&Det.&Und.&All&Det.&Und.&All&Det.&Und.&All&Det.&Und.&All&Det.&Und.\\
\hline
       1&$Starbursts                           $&      14&       6&       8&0.46&0.44&0.47&41.41&41.26&41.49&\textbf{9960 $\pm$ 500}&\textbf{22400 $\pm$ 1200}&\textbf{630 $\pm$ 110}& 736.8&1592.9&  94.7&11.0&11.3&10.1\\
       2&$z<1~L_X=10^{41-42}                   $&      25&       3&      22&0.43&0.29&0.45&41.60&41.12&41.64&\textbf{1540 $\pm$ 160}&\textbf{10070 $\pm$ 550}&380 $\pm$ 170& 235.7&1183.5& 106.4&10.4&10.8&$<$10.5\\
       3&$z>3                                  $&      12&       0&      12&3.52&-&3.52&44.31&-&44.31&280 $\pm$ 180&-&280 $\pm$ 180& 127.5&-& 127.5&-&-&-\\
       4&$BLAGN                                $&      34&       6&      28&1.54&1.02&1.65&43.94&43.74&43.97&\textbf{2050 $\pm$ 150}&\textbf{10050 $\pm$ 450}&330 $\pm$ 160& 419.3& 957.3& 304.1&12.0&12.3&$<$11.9\\
       5&$z=0.5-1~L_X=10^{42-44}~N_H<10^{22}   $&      32&       3&      29&0.74&0.72&0.74&43.10&42.88&43.12&\textbf{2180 $\pm$ 170}&\textbf{14200 $\pm$ 830}&\textbf{940 $\pm$ 170}& 301.6& 820.2& 247.9&11.2&11.7&10.9\\
       6&$z=1-2~L_X=10^{42-44}~N_H<10^{22}     $&      19&       2&      17&1.38&1.03&1.42&43.39&43.61&43.35&\textbf{1230 $\pm$ 160}&\textbf{8080 $\pm$ 540}&420 $\pm$ 170& 265.0&1156.6& 160.1&11.7&12.2&$<$11.7\\
       7&$z=0.5-1~L_X=10^{42-44}~N_H=10^{22-23}$&      31&       3&      28&0.75&0.67&0.76&42.99&43.20&42.96&\textbf{750 $\pm$ 140}&\textbf{4700 $\pm$ 380}&330 $\pm$ 150& 180.0& 494.0& 146.3&10.8&11.4&$<$10.9\\
       8&$z=1-2~L_X=10^{42-44}~N_H=10^{22-23}  $&      32&       2&      30&1.38&1.32&1.39&43.35&43.59&43.33&\textbf{680 $\pm$ 140}&\textbf{4300 $\pm$ 330}&440 $\pm$ 150& 174.4& 480.3& 154.0&11.4&12.3&$<$11.7\\
       9&$z=0.5-1~L_X=10^{42-44}~N_H<10^{23}   $&      63&       6&      57&0.74&0.69&0.75&43.05&43.07&43.05&\textbf{1440 $\pm$ 110}&\textbf{9450 $\pm$ 460}&\textbf{600 $\pm$ 110}& 241.7& 657.1& 198.0&11.0&11.5&10.7\\
      10&$z=1-2~L_X=10^{42-44}~N_H<10^{23}     $&      51&       4&      47&1.38&1.18&1.40&43.36&43.60&43.34&\textbf{920 $\pm$ 110}&\textbf{6190 $\pm$ 320}&\textbf{470 $\pm$ 120}& 208.1& 818.5& 156.2&11.5&12.3&11.2\\
      11&$z=0.5-1~L_X=10^{42-43}~N_H<10^{23}   $&      44&       3&      41&0.74&0.58&0.75&42.63&42.65&42.63&\textbf{1390 $\pm$ 130}&\textbf{12850 $\pm$ 820}&\textbf{550 $\pm$ 130}& 167.8& 688.3& 129.7&11.0&11.4&10.7\\
      12&$z=1-2~L_X=10^{42-43}~N_H<10^{23}     $&      19&       1&      18&1.35&1.03&1.37&42.68&42.49&42.69&\textbf{1110 $\pm$ 170}&\textbf{8150 $\pm$ 750}&\textbf{720 $\pm$ 170}& 153.0& 555.6& 130.6&11.7&11.9&11.4\\
      13&$z=0.5-1~L_X=10^{43-44}~N_H<10^{23}   $&      19&       3&      16&0.77&0.81&0.76&43.43&43.28&43.46&\textbf{1510 $\pm$ 200}&\textbf{6060 $\pm$ 410}&660 $\pm$ 230& 413.1& 625.9& 373.2&11.1&11.8&$<$11.2\\
      14&$z=1-2~L_X=10^{43-44}~N_H<10^{23}     $&      32&       3&      29&1.40&1.23&1.42&43.53&43.72&43.51&\textbf{870 $\pm$ 140}&\textbf{5540 $\pm$ 340}&390 $\pm$ 150& 240.9& 906.1& 172.1&11.5&12.3&$<$11.7\\
      15&$S_{24}=50{\mu}Jy-150{\mu}Jy          $&      68&       0&      68&1.40&-&1.40&43.73&-&43.73&260 $\pm$ 90&-&260 $\pm$ 90& 113.2&-& 113.2&$<$11.4&-&$<$11.4\\
      16&$S_{24}=150{\mu}Jy-500{\mu}Jy         $&      68&      10&      58&1.37&1.00&1.44&43.75&43.52&43.78&\textbf{1263 $\pm$ 97}&\textbf{4420 $\pm$ 180}&\textbf{720 $\pm$ 110}& 285.9& 456.9& 256.4&11.7&11.9&11.5\\
      17&$S_{24}\geq500{\mu}Jy                 $&      20&      13&       7&1.10&1.01&1.27&44.02&44.01&44.04&\textbf{7040 $\pm$ 230}&\textbf{10120 $\pm$ 300}&\textbf{1330 $\pm$ 320}&1113.4&1199.7& 953.0&12.2&12.3&11.6\\
      18&$S_{24}\geq50{\mu}Jy                  $&     156&      23&     133&1.35&1.01&1.41&43.79&43.86&43.77&\textbf{1570 $\pm$ 66}&\textbf{7640 $\pm$ 190}&\textbf{520 $\pm$ 70}& 316.7& 876.8& 219.9&11.7&12.1&11.3\\
\hline
\end{tabular}

\label{Stack_Info}
}\end{center} {\sc Notes}: (1) Stack index. (2) Stack description. (3)
The numbers of CDF-S AGNs/SBs in each stack. (4) Mean redshift of the
sources in each stack. (5) Logarithm of the mean, absorption corrected
X-ray luminosity of the sources in each stack. (7) Stacked 70~\micron\
flux density per source. (8) Stacked 24~\micron\ flux density per
source. (9) Infrared luminosity per source (over the 8-1000~\micron\
interval), derived from the 70~\micron\ flux density.  Upper limits
include the uncertainty on the $S_{70}$--$L_{\rm IR}$ correction.  We
do not calculate $L_{\rm IR}$ for stack (4), $z>3$ AGN, as the mean
redshift is beyond the range of the tracks derived from the BAT/IRS
sample.
\end{table}
\end{landscape}

\begin{table*}
  \caption{Average properties of the CDF-S X-ray AGNs in the 70~\micron\  \textit{undetected}, `Restricted' sample (see \S\ref{Dat_Xray}).  $N_{all}$ is the number of AGNs detected in each bin and $N_{det}$ is the number of these that are detected at 24~\micron. $\overline{z}$ is the mean redshift of the AGNs in each subsample and ${\rm log}(\overline{L_{\rm X}})$ and ${\rm log}(\widetilde{L_{\rm X}})$ indicate their mean and median \Lx.  We provide the mean \lratio\ ratios derived from the 70~\micron\ and 24~\micron\ stacks (assuming the Average AGN and Average QSO SEDs for the \Lx$=10^{42-43}$~\ergs\ and \Lx$=10^{43-44}$~\ergs\ subsamples, respectively) as well as the mean and median \lratio\ ratios derived from the individual 24~\micron\ flux densities.  Finally, we provide the mean and median \lratio\ derived from the individual 24~\micron\ flux densities assuming the two BAT/IRAS tracks that most closely match the \fratio\ flux ratios of the \Lx$=10^{42-43}$~\ergs\ and \Lx$=10^{43-44}$~\ergs\ subsamples at both $z=0.5-1$ and $z=1-2$: NGC~1275 and Mrk~290, respectively.}  
  \begin{tabular}{@{}lccccccccccc@{}}
\hline
\hline
&&&&&&\multicolumn{2}{c}{\it Stacks}&\multicolumn{2}{c}{\it Avg. AGN/QSO}&\multicolumn{2}{c}{\it BAT/IRAS closest match}\\
&$N_{all}$&$N_{det}$&$\overline{z}$&${\rm log}(\overline{L_X})$&${\rm log}(\widetilde{L_X})$&$\frac{\overline{L_{IR,70}}}{\overline{L_X}}$&$\frac{\overline{L_{IR,24}}}{\overline{L_X}}$&$\frac{\overline{L_{IR,24}}}{\overline{L_X}}$&$\frac{\widetilde{L_{IR,24}}}{\widetilde{L_X}}$&$\frac{\overline{L_{IR,24}}}{\overline{L_X}}$&$\frac{\widetilde{L_{IR,24}}}{\widetilde{L_X}}$\\
\hline
$z=0.5-1, L_{X}=10^{42-43}{\rm erg/s}$&  41&  28&0.7&42.6&42.6& 42.7& 26.2& 24.9& 18.0& 28.1& 20.3\\
$z=1-2, L_{X}=10^{42-43}{\rm erg/s}$&  19&  10&1.4&42.7&42.6&201.2&133.3&112.6& 80.2&164.2&100.2\\
$z=0.5-1, L_{X}=10^{43-44}{\rm erg/s}$&  16&  15&0.8&43.5&43.4&$<$ 17.4&  7.7&  6.4&  5.3&  8.9&  7.7\\
$z=1-2, L_{X}=10^{43-44}{\rm erg/s}$&  30&  19&1.4&43.5&43.4&$<$ 35.1& 13.9& 11.4&  7.8& 19.9& 14.1\\
\hline
\end{tabular}

  \label{F24_Info}
\end{table*}

\subsection{24~\micron\ Properties}
\label{Dis_24um}

The results on the $L_{\rm IR}/L_{\rm X}$ ratios in \S\ref{Ana_Stack}
were based on stacking analyses, which could be biased by a few bright
70~\micron\ undetected sources. Since the majority of the CDF-S X-ray
AGNs are individually detected at 24~\micron, we can perform a
complementary test of our results that does not rely significantly on
stacking analyses. The advantage of this approach is that we can
assess the range of $L_{\rm IR}$ for the X-ray AGNs, although there
can be considerable uncertainty in the conversion between 24~\micron\
flux density and $L_{\rm IR}$ for individual objects; see
\Fig{Fluxes}. However, as we show in \Fig{Fluxes}, provided the
average SED is not predominantly starburst dominated, then we can
predict the average $L_{\rm IR}/L_{\rm X}$ ratio, on the basis of
  the 24 \micron\ data, to within a factor of $\approx$~3; on the
basis of \Fig{Ratios}, the average SEDs are AGN dominated, and would
be increasingly so if the stacked 70~\micron\ flux is dominated by
bright undetected sources. Furthermore, so long as the average SEDs of
the $z=$~0.5--1 AGNs are the same as the $z=$~1--2 AGNs, then we can
accurately assess the relative change in $L_{\rm IR}/L_{\rm X}$
between these redshifts; we provide evidence in \S4.1 that this does
appear to be the case. To convert from 24~\micron\ flux densities to
$L_{\rm IR}$ we use the same SEDs as assumed when calculating \Lir\
from the stacked 70~\micron\ flux densities (i.e., Average AGN and
Average QSO for the \Lx$=10^{42-43}$~\ergs\ and $10^{43-44}$~\ergs\
AGNs, respectively; see \S\ref{Ana_Stack_Lum}).  For those AGNs that
are undetected at 24~\micron\ we calculated upper limits on $L_{\rm
  IR}$ assuming $S_{24}=50$~\mujy; however, we note that this
assumption has a small effect on our average results (i.e.,\
$\overline{L_{\rm IR}}/\overline{L_{\rm X}}$ only changes by
$\approx$~35\% even if we assume extreme upper limits of
$S_{24}=0$~\mujy). The average \lratio\ ratios derived from the
24~\micron\ fluxes are given in Table~1.

We include in \Fig{LIRLX_vs_z} the individual \lratio\ ratios
calculated using the \Lir\ values derived from 24~\micron\ fluxes of
the 70~\micron\ {\it undetected} AGNs in the `Restricted' sample. For
both the \Lx$=10^{42-43}$\ergs\ and \Lx$=10^{43-44}$\ergs\ AGNs the
mean \lratio\ ratios increase with redshift by factors of $\approx$4.5
and $\approx$1.8, respectively.  These results are in qualitative
  agreement with those obtained from the 70~\micron\ stacks and
indicates that the stacks are not dominated by a few extreme sources;
these results are also confirmed by the median \lratio\ ratios
presented in \Tab{F24_Info}. However, we note that there is a factor
of $\approx$2 offset between the \lratio\ ratios derived from the
70~\micron\ stacks and individual 24~\micron\ flux densities. This is
caused by the difference between the \fratio\ flux ratios of the
70~\micron\ {\it undetected} stacks and the assumed SEDs (see
\Fig{Ratios}) and disappears if we assume the SEDs of the two
individual BAT/IRS AGNs that most closely reproduce (at both redshift
bins) the \fratio\ flux ratios of the 70~\micron\ {\it undetected}
\Lx$=10^{42-43}$~\ergs\ and $10^{43-44}$~\ergs\ stacks: NGC~1275 and
Mrk~290, respectively.

%
\section{Discussion}
%

We have provided multiple lines of evidence for an increase in the
\lratio\ ratio for AGNs over the redshift range $z=0-2$, with
\Lx$=10^{42-43}$~\ergs\ AGNs $\approx$5 times more infrared luminous
at $z=1-2$ than at $z=0.5-1$ (and $\approx20$ times more infrared
luminous than at $z\approx 0$).  The evidence for an increase in \Lir\
for \Lx$=10^{43-44}$~\ergs\ AGNs is less conclusive: a factor of
$\approx$2 between $z=0.5-1$ and $z=1-2$ and no change from $z\approx
0$ and $z=1-2$.

These results provide new insight into
the production of IR emission from distant AGNs and lead to a number
of questions, which we address below.

\subsection{What is driving the increase in $L_{\rm IR}$?}
\label{Dis_SB_AGN}
The \lratio\ ratio for the \Lx$=10^{42-43}$~\ergs\ AGNs at $z=1-2$ is
higher than that found for lower redshift X-ray AGNs; see
\Tab{F24_Info}. However, the \lratio\ ratio for the
\Lx$=10^{42-43}$~\ergs\ AGNs is broadly consistent with $z\approx 2$
submillimeter emitting galaxies (SMGs) hosting AGN activity
($L_{\rm FIR}/L_{\rm X}$$\approx$250; \citealt{Alexander05}).  On the
basis of sensitive \Spitzer -IRS spectroscopy, the large $L_{\rm
  FIR}/L_{\rm X}$ ratio from AGN-hosting SMGs appears to be due to
intense star-formation activity, with an average contribution to
$L_{\rm FIR}$ from AGN activity of $\approx$10\% (e.g.,
\citealt{Pope08, Menendez-Delmestre07, Menendez-Delmestre09}). Can the
increase in the average \lratio\ ratio of the $z=1-2$ AGNs also be due
to increased star-formation activity?

There is no clear relative offset between the stacked \fratio\ flux
ratio and the AGN-dominated track for the \Lx$=10^{42-43}$~\ergs\ AGNs
over the redshift range $z=0.5-2$, appearing to suggest that the
relative AGN/SB contribution to $L_{\rm IR}$ has not changed with
redshift. If we derive the relative AGN/SB contribution to the stacked
\fratio\ using the average AGN and SB tracks, then we estimate
$\approx$60\% and $\approx$70\% of the AGNs with
\Lx$=10^{42-43}$~\ergs\ have AGN-dominated SEDs at $z=0.5-1$ and
$z=1-2$, respectively; by comparison, the stacked \fratio\ of the
\Lx$=10^{43-44}$~\ergs\ AGNs suggest $\approx$100\% have AGN-dominated
SEDs over the full redshift range. However, since it is difficult to
unambiguously determine the relative AGN/SB contribution to $L_{\rm
  IR}$ from the \fratio\ flux ratios at $z>1.5$ (see
Fig.~\ref{Ratios}), this result should be considered
tentative.\footnote{We note that we also find a similar fraction
  ($\approx30$\%) of \Lx$=10^{42-43}$~\ergs\ AGNs at $z=1-2$ have IRAC
  3.6--8.0~\micron\ colours suggesting AGN-dominated SEDs (based on
  \citealt{Stern05}) as that found for \Lx$=10^{42-43}$~\ergs\ AGNs at
  $z=0.5-1$. This also suggests that the relative AGN/SB contribution
  for \Lx$=10^{42-43}$~\ergs\ AGNs is constant with redshift but we
  caution that this result is based on shorter-wavelength data than
  used to derive \Lir.} Therefore, we explore below the implication of
our results assuming both an increase in star formation and AGN
activity.

If the observed increase in \Lir\ is attributed to star-formation it
would imply a significantly higher (i.e., a factor of
$\approx$5) ratio between star-formation and black hole growth at
$z=1-2$ compared to $z=0.5-1$, and a factor of $\approx 13$
  increase between $z\approx 0$ and $z=1-2$.  Whether this has
significant implications for the black-hole--bulge mass relationship
depends on the location of the star-formation in the host galaxy and
on the AGN fraction (i.e.,\ the duty cycle of black-hole growth). For
example, using Eqn. 4 of \cite{Kennicutt98} to derive the
star-formation rate from \Lir\ and deriving the mass accretion rate
from \Lx, we estimate that the ratio between the average
star-formation and mass accretion rate is $\approx$500 in the $z=1-2$
AGNs, which is consistent with the black-hole to bulge mass ratio
observed in the local universe (i.e., $\approx$800; \citealt{Mclure02,
  Marconi03}). This would appear to suggest a closer relationship
between black-hole and stellar growth at $z=1-2$ than found at
$z=0.5-1$, where the average star-formation and mass accretion rate
would be $\approx$100 (and $\approx$30 at $z\approx 0$), based
on the same assumptions as above. However, these different results
could be reconciled if the AGN fraction is higher at higher redshifts
or if the majority of the star formation at higher redshifts is
occurring in the galaxy disk rather than the galaxy bulge, or
vice-versa.

If, however, the increase in the \lratio\ ratio is due to the AGN, it
would imply that a larger fraction of the intrinsic emission from the
accretion disk is reprocessed by dust.  This increase in the \lratio\
ratio may therefore be due to larger AGN dust covering factors at
higher redshifts; there is tentative evidence that this is indeed the
case from measurements of the obscured to unobscured ratios of AGNs in
deep X-ray surveys (e.g. \citealt{LaFranca05, Hasinger08}). Results
presented in \cite{Hasinger08} predict a factor of $\approx$2 increase
in the ratio of obscured-to-unobscured AGNs between $z=0$ and $z=1-2$,
significantly lower than what we measure here.  However, this is
based on a sample of AGNs covering a broader range of \Lx\ than
focused on here and there is evidence that any increase in the dust
covering fraction will be weaker in more X-ray luminous AGNs
(e.g. \citealt{Ueda03,Akylas06,Treister08}; and supported by the
reduced \lratio\ ratios of the \Lx$=10^{43-44}$~\ergs\ AGNs reported
here). Therefore, on the basis of the X-ray survey results, it is
plausible that at least some of the increase in \lratio\ ratio is due
to a larger dust covering factor at $z\approx1-2$ than seen at lower
redshifts. An increase in dust-covering factor with redshift is
predicted by a number of theoretical models, which suggest that AGNs
undergo early growth during a hidden phase before expelling their
obscuring gas and dust, revealing a luminous, unobscured quasar
(\citealt{Silk98,Springel05,Hopkins06}).

More direct constraints on the origin of the increase in the \lratio\
ratio will be placed using the \Herschel\ {\it Space Observatory} (see
\S\ref{Dis_Herschel}) and {\it Spitzer}-IRS, for the fraction of X-ray
AGNs that have {\it Spitzer}-IRS data (J.~R.~Mullaney, in
preparation). 

\begin{figure}
\includegraphics[angle=0,width=85mm]{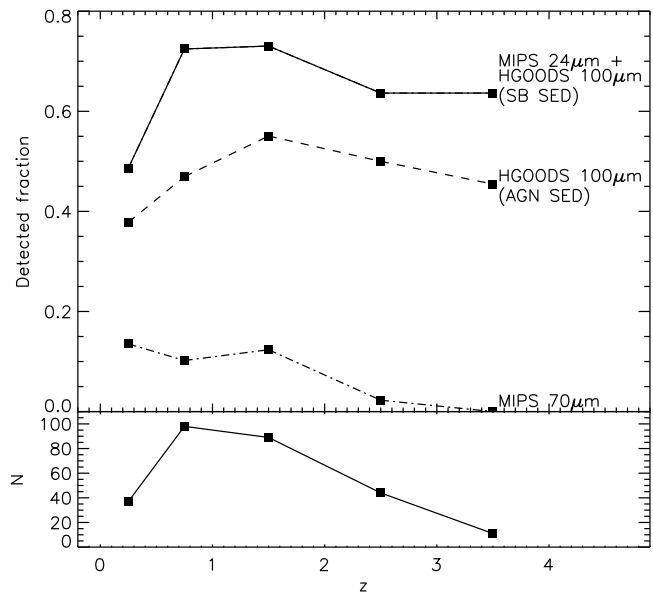}
\caption{\comment{LABEL:H\_Detect; }\textit{Top:} Predicted fractions
  of CDF-S X-ray AGNs detected in the upcoming HGOODS deep infrared
  survey to be undertaken by the \textit{Herschel Space Observatory}.
  The expected observed 100~\micron\ flux densities are calculated by
  extrapolating the \Spitzer-MIPS 24~\micron\ flux densities along the
  Average SB-dominated and Average AGN-dominated SEDs, derived from
  the BAT/IRS sample of AGNs.  If all the X-ray AGNs were to have
    SB-dominated IR SEDs, we would expect to detect {\it at least} as high a
    fraction at 100~\micron\ as we currently detect at 24~\micron.
    Also shown for comparison are the fractions of CDF-S AGNs detected
    by MIPS at 70~\micron\ in each redshift bin \textit{Bottom:} The
    number of CDF-S X-ray AGNs in each redshift bin.}
\label{H_Detect}
\end{figure}

\begin{figure}
\includegraphics[angle=0,width=85mm]{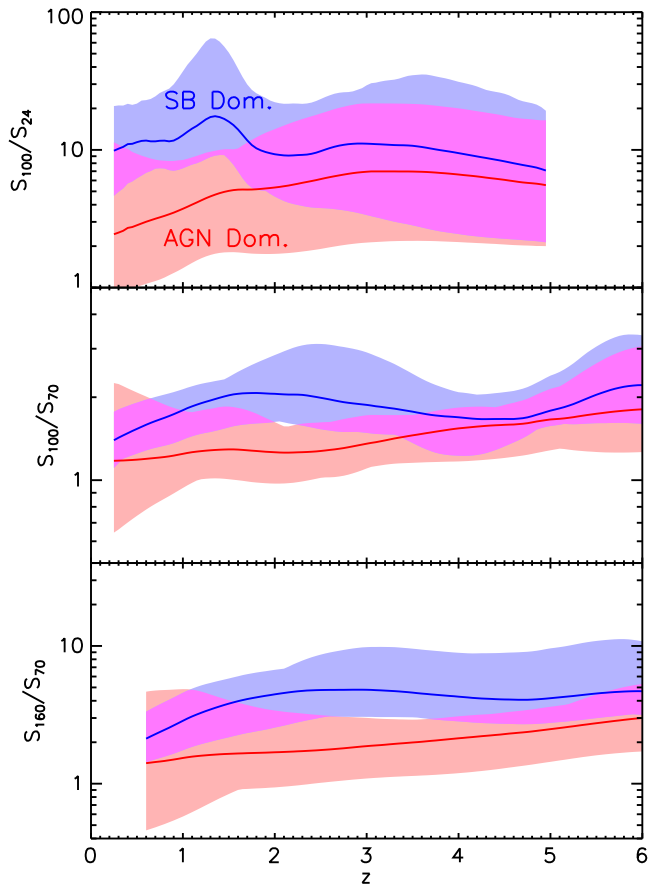}
\caption{\comment{LABEL:H\_Tracks; } AGN/SB diagnostic powers of the
  PACS filters on board \textit{Herschel Space Observatory}.  The
  shaded regions indicate the range of expected flux ratios of the
  AGNs in the BAT/IRS sample at the redshifts shown. Solid lines
  indicate the expected flux ratios of the BAT/IRS Average AGN and
  Average SB.}
\label{H_Tracks}
\end{figure}

\subsection{What is the contribution of AGNs to the cosmic IR
  background?}
\label{Dis_Contrib}
We can use the results of our stacking analysis to place constraints
on the AGN contribution to the 70~\micron\ background. If we stack all
of the 251 X-ray AGNs, irrespective of whether they are individually
detected at 70~\micron\ we obtain an average 70~\micron\ flux density
of 1040$\pm$80~\mujy, and therefore an integrated 70~\micron\ flux
density of 260$\pm$20~mJy, over the 391.3~arcmin$^2$ of the CDF-S
field.  On the basis of the analysis of the IR background in both
GOODS fields and the Lockman Hole (e.g., \citealt{Dole06}), this
corresponds to $\gtrsim$5\% of the average resolved 70~\micron\
background.  This constraint is a lower limit since (1) the X-ray
observations will not have identified the most heavily obscured
luminous AGNs in this field, (2) the 70~\micron\ field is too small to
include X-ray bright AGNs, and (3) the X-ray observations are not
sensitive to the lowest-luminosity AGNs.

On the basis of \cite{Tozzi06}, we would predict $\approx$80\% of
\Lx$>10^{41}$\ergs, \Nh$<10^{23}$ AGNs in the CDF-S (i.e., to
$z\approx$5) to be X-ray undetected. If we assume that they have the
same IR SED as the X-ray detected AGNs then they increase our estimate
of the resolved 70~\micron\ background by a factor of five to
  $\gtrsim$25\%; from a study of nearby sources, \cite{Lutz04} finds
that the IR emission of AGNs is not significantly depressed in the
most heavily obscured sources (we confirm that this is also the case
for the BAT/IRS sample). Since X-ray surveys are insensitive to the
most heavily obscured AGNs (i.e.,\ Compton-thick sources with
\Nh$>10^{24}$~cm$^{-2}$), the overall contribution to the 70~\micron\
background could potentially double to $\gtrsim10$\% (e.g.,\
\citealt{Daddi07, Alexander08}).

To estimate the contribution to the 70~\micron\ background from bright
AGNs that are too rare to lie in the small CDF-S field would require
the same analyses as performed here to be made on a shallower survey
that covers a larger area of the sky.  As it stands, the most accurate
70~\micron\ analysis in a larger field is that of \cite{Papovich07} of
the E-CDFS.  They find that bright AGNs (i.e., \Lx$>10^{44}$\ergs)
contribute only 7\e{-4}~\MJysr\ to the IR background at 70~\micron,
corresponding to an insignificant fraction ($\ll1$\%) of the total IR
background at this wavelength.  However, the E-CDFS field is not a
great deal larger than the CDF-S field and a full study of the
contribution of X-ray bright AGNs to the IR background will require
the analysis of the much larger COSMOS\footnote{URL:
  http://cosmos.astro.caltech.edu/} and/or Bootes\footnote{URL:
  http://www.lsstmail.org/noao/noaodeep/} fields at 70~\micron.

\subsection{What advances may we expect for deep surveys with \Herschel?}
\label{Dis_Herschel}
Our study of the FIR properties of X-ray detected AGNs allows us to
place constraints on the properties of AGNs detected in deep {\it
  Herschel} fields at $>70$~\micron.  We anticipate that the improved
sensitivity of the \Herschel-PACS instrument over \Spitzer-MIPS will
result in a significant increase in the fractions of X-ray AGNs that
will be detected at FIR wavelengths, which will provide more direct
insight into the processes driving the increase in \Lir\ at $z=1-2$.
Depending on the assumed SED we predict that approximately
  45-60\%, 50-70\%, 50-65\% and 45-65\% of the CDF-S X-ray sources at
$z<1$, $z = 1-2, 2-3$ and $3-4$, respectively, will be detected at
100~\micron\ in the proposed ultradeep HGOODS survey (limiting flux =
0.6~\mjy; see \Fig{H_Detect}).  These estimated detection levels
  are calculated by extrapolating the 24~\micron\ flux densities along
  the Average SB-dominated and Average AGN-dominated SEDs (derived from
  the BAT/IRS sample).

In \Fig{H_Tracks} we present the results of passing the IR spectra of
the BAT/IRS sample through the various PACS filter response curves.
Our analysis shows that, although the $S_{100}/S_{24}$ flux ratio will
not push our constraints on the relative contributions from AGN
activity and star formation to higher redshifts (due to the influence
of the MIR spectral features that shift into the observed 24~\micron\
band at high redshifts), the $S_{100}/S_{70}$ and $S_{160}/S_{70}$
flux ratios will ``take over'' and provide this information out to a
redshift of $\sim 5$ (i.e., the redshift of the most distant AGNs
currently detected in the deepest X-ray surveys).  Deep, infrared
  observations undertaken by \Herschel\ will therefore provide a
  method of identifying a significant proportion of the long sought
  after population of Compton-thick AGNs out to $z\approx$6.

\section{Summary}
We have investigated the MIR and FIR properties of X-ray detected AGNs
and SBs in the CDF-S using deep 24~\micron\ and 70~\micron\
observations undertaken by the \Spitzer\ {\it Space Telescope} as part
of the GOODS and FIDEL legacy surveys.  Out of the 266 X-ray AGNs/SBs,
30 ($\approx$11\%) and 172 ($\approx$65\%) have counterparts at
70~\micron\ and 24~\micron\, respectively, with a bias towards the
detection of SBs and BLAGNs at 70~\micron.  As the majority of CDF-S
X-ray AGNs are undetected at 70~\micron\ we rely on stacking analysis
to measure their average MIR and FIR properties.  We compare the IR
properties of the CDF-S AGNs/SBs with those of a sample of local AGNs
that have X-ray properties (i.e., \Lx\ and \Nh) covering the same
range as the CDF-S AGNs (i.e., the BAT/IRS sample).  In the following
points we summarise the main conclusions of this study:

\begin{enumerate}
\item \label{Brighter} We find strong evidence at both 24~\micron\ and
  70~\micron\ that the average infrared luminosity of
  \Lx$=10^{42-43}$~\ergs\ AGNs at $z=1-2$ is significantly higher than
  those at $z=0.5-1$ and $z\approx0$ (i.e., by a factor of $\approx5$
  and $\approx20$, respectively).  This difference corresponds to a
  factor of $4.7^{+10.2}_{-2.0}$ and $12.7^{+7.1}_{-2.6}$ increase in
  \Lir/\Lx, respectively.  This large increase in \Lir\ is not,
  however, seen for AGNs with higher X-ray luminosities (i.e.,
  \Lx$=10^{43-44}$~\ergs).  We therefore argue that deep IR
  observations can be used as a reliable method to identify
  intrinsically bright (i.e.., \Lx$=10^{43-44}$~\ergs) Compton-thick
  AGNs, but urge caution when using this ratio to locate less luminous
  obscured AGNs (i.e., $L_{\rm X}\lesssim 10^{43}$~\ergs), unless
  other indicators of AGN activity are present.
\item Due to the low numbers of AGNs detected at 70~\micron\ we are
  unable to establish what process is driving this increase in the
  average \Lir; however, both increased star-formation and/or
  increased AGN dust covering factors are likely candidates.  If the
  former, then the $z=1-2$ epoch may represent a period of rapid
  growth of the bulge to black hole mass ratio.  However, there is
  tentative evidence from X-ray observations that the dust covering
  fraction is, indeed, higher at large redshifts.  We predict that
  forthcoming deep surveys to be carried out by the \Herschel\ {\it
    Space Observatory} will enable us to resolve what process is
  driving the increase in \Lir.  See sections \ref{Ana_Stack_Lum},
  \ref{Dis_24um}, \ref{Dis_SB_AGN}.
\item On average, more X-ray luminous CDF-S AGNs have lower \fratio\
  flux ratios.  Based on the infrared properties of a sample of local
  AGNs with similar X-ray properties (i.e., \Lx\ and \Nh), we conclude
  that more X-ray luminous CDF-S AGNs have IR SEDs that are more
  AGN-dominated (rather than SB dominated).  See section
  \ref{Ana_Stack_Ratio}.
\item Despite measuring a large increase in \Lir\ among $z=1-2$ AGNs
  (see point \ref{Brighter} above), we find that the X-ray detected
  AGNs in the CDF-S contribute only $\approx$5\% of the 70~\micron\
  background.  However, if we extrapolate this fraction to take into
  account those $\approx$80\% of AGNs that are undetected in X-rays, we
  estimate that $\approx$25\% of the 70~\micron\ background is
  attributable to AGNs.  See section \ref{Dis_Contrib}.
\item We anticipate that the undertaking of deep, FIR surveys by the
  \Herschel\ {\it Space Observatory} will allow us to detect between
  40\% and 75\% of the X-ray detected AGNs in the 1~Ms CDF-S,
  depending on whether the infrared SEDs are predominantly AGN or
  SB-dominated, respectively.  Furthermore, the FIR diagnostics used
  in our analysis can be directly applied to the PACS filters on board
  \Herschel\ to enable us to discriminate between AGNs with
  SB-dominated or AGN-dominated infrared SEDs. See section
  \ref{Dis_Herschel}.
\end{enumerate}

\section*{acknowledgements}
We would like to thank David Elbaz for his useful comments on the
paper and for providing us with the PACS filter response curves.  We
would also like to thank Chris Done for her useful comments on the
bolometric luminosities of X-ray detected AGNs.  Furthermore, we would
like to thank the anonymous referee for their comments, particularly
those concerning the X-ray to IR matching procedure. This work is
based (in part) on observations made with the {\it Spitzer Space
  Telescope} and has made use of the NASA/ IPAC Infrared Science
Archive, which are both operated by the Jet Propulsion Laboratory,
California Institute of Technology under a contract with NASA. Support
for this work was provided by NASA through an award issued by
JPL/Caltech. We gratefully acknowledge support from the Leverhulme
Trust (JRM; DMA) and the Royal Society (DMA).

\bsp

\appendix
\section{The \Swift-BAT Comparison Sample}
\label{BAT_sample}
To aid in the interpretation of the {\it Spitzer} MIPS data used to
characterise the CDF-S X-ray sample, we selected a local sample of
AGNs from the {\it Swift}-BAT survey with {\it Spitzer}-IRS
low-resolution spectroscopy (5.2--38~\micron). The {\it Spitzer}-IRS
data allows us to accurately distinguish between AGNs with
AGN-dominated and SB-dominated IR SEDs on the basis of the MIR
spectral features in the {\it Swift}-BAT AGNs. The {\it Swift}-BAT
survey provides the ideal local AGN comparison sample since the
sensitivity of the BAT telescope to hard X-ray photons (14--195~keV)
provides an almost absorption-independent selection of AGNs (to
\Nh$\approx10^{24}$~\cs). Indeed, the range of X-ray luminosities and
absorbing column densities of the {\it Swift}-BAT AGNs are comparable
to those of the CDF-S AGNs (see \citealt{Tueller08} and
\citealt{Winter09}); see \S2.2.1.

\begin{figure*}
\includegraphics[angle=90,width=170mm]{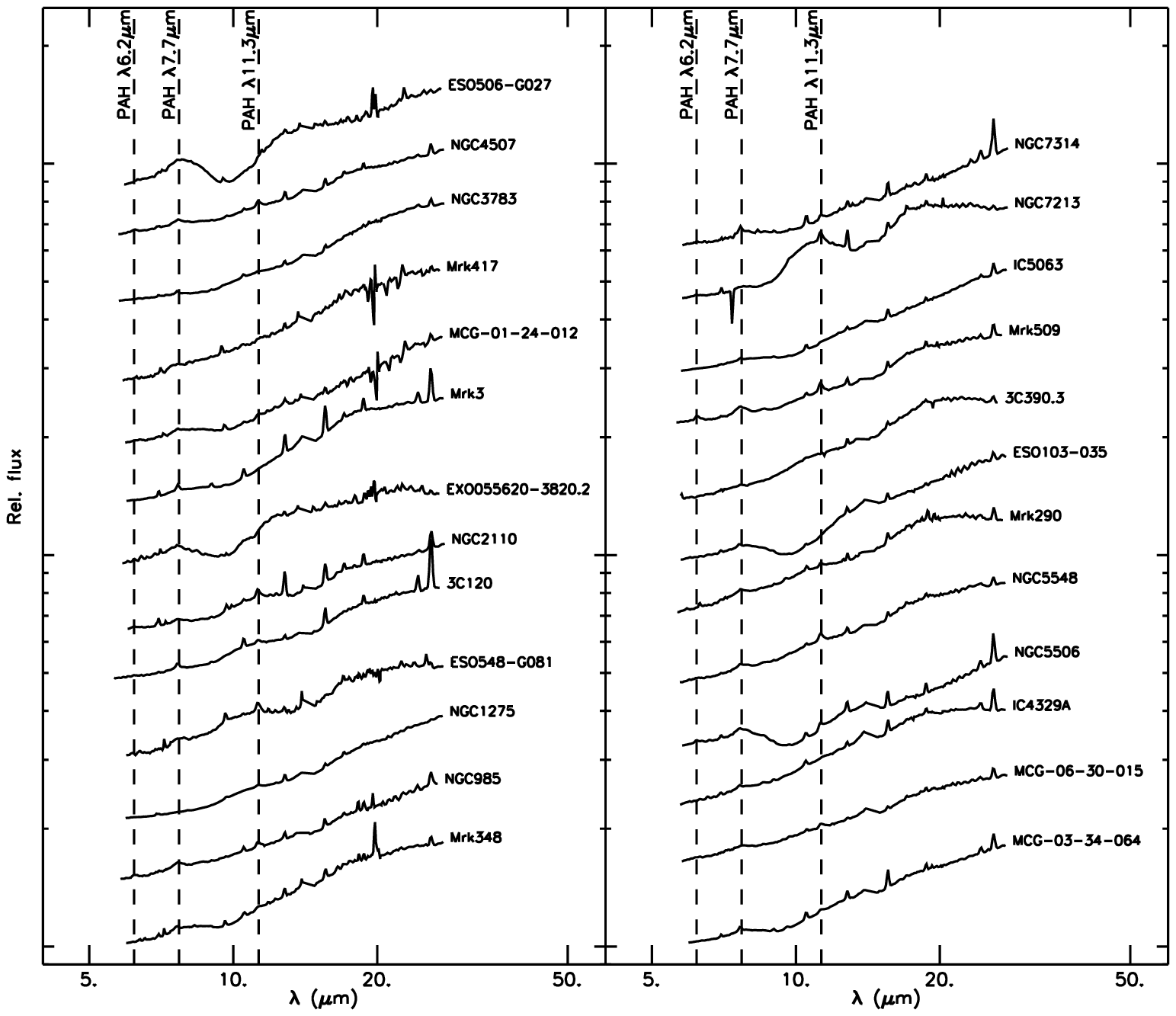}
\caption{IRS spectra of the 25 BAT/IRS AGNs that we classify as having
  AGN-dominated IR SEDs on the basis of the lack of a prominent PAH
  feature at 6.2~\micron.  Also indicated are the PAH features at
  7.7~\micron\ and 11.4~\micron\, although these were not used to
  distinguish between AGN and SB dominated systems.}
\label{W09_AGN}
\end{figure*}

Thirty-six of the 104 {\it Swift}-BAT AGNs with good-quality
$L_{2-10~keV}$ and \Nh\ constraints from \cite{Winter09} or
\cite{Bassani99} have both Short-Low (SL; 3.6" x 136.0";
$R\sim$~60--127) and Long-Low (LL; 10.5" x 360"; $R\sim$~57--126) {\it
  Spitzer}-IRS spectroscopy, providing full coverage at
5.2--38~\micron.  All 36 BAT/IRS AGNs were observed in staring-mode
with two nod positions, which are required for background
subtraction. Basic Calibrated Data (BCD) images were combined and
cleaned as described in Goulding \& Alexander (2009). Differing nod
positions were subtracted from each other to produce background
subtracted images before extracting the spectra using {\sc
  spice}\footnote{URL:
    http://ssc.spitzer.caltech.edu/postbcd/spice.html}. The objects
were sufficiently bright and the observations were sufficiently short
that there was no significant effect from latent charge build-up on
the detector. Long-slit orders were clipped using the wavelength trim
ranges given in Table 5.1 of the {\it Spitzer}-IRS Observers
manual\footnote{The {\it Spitzer}-IRS Observers Manual is available at
  http://ssc.spitzer.caltech.edu/irs/dh/}. The continuum of each
source in every order was fitted using a first or second-order
polynomial. The spectra were then matched to give a single continuous
spectrum for each source. Flux calibration was carried out using the
latest available Spitzer-IRS calibration files (version 17.2) and is
largely consistent with archival \IRAS\ 25~\micron\ flux densities (to
within a factor of 2.5 in $\approx$97\% of cases).  The BAT/IRS
  AGNs have absorption corrected 2-10~keV luminosities spanning
  4.3\e{41}~\ergs\ to 1.9\e{44}~\ergs\ and absorbing column densities
  in the range 2.5\e{20}~\cs\ to 2.2\e{24}~\cs\ (i.e., roughly the
  same as those CDF-S X-ray AGN/SBs with \Lx\ and \Nh\ measurements
  published in \citealt{Tozzi06}).

\begin{figure*}
\includegraphics[angle=0,width=85mm]{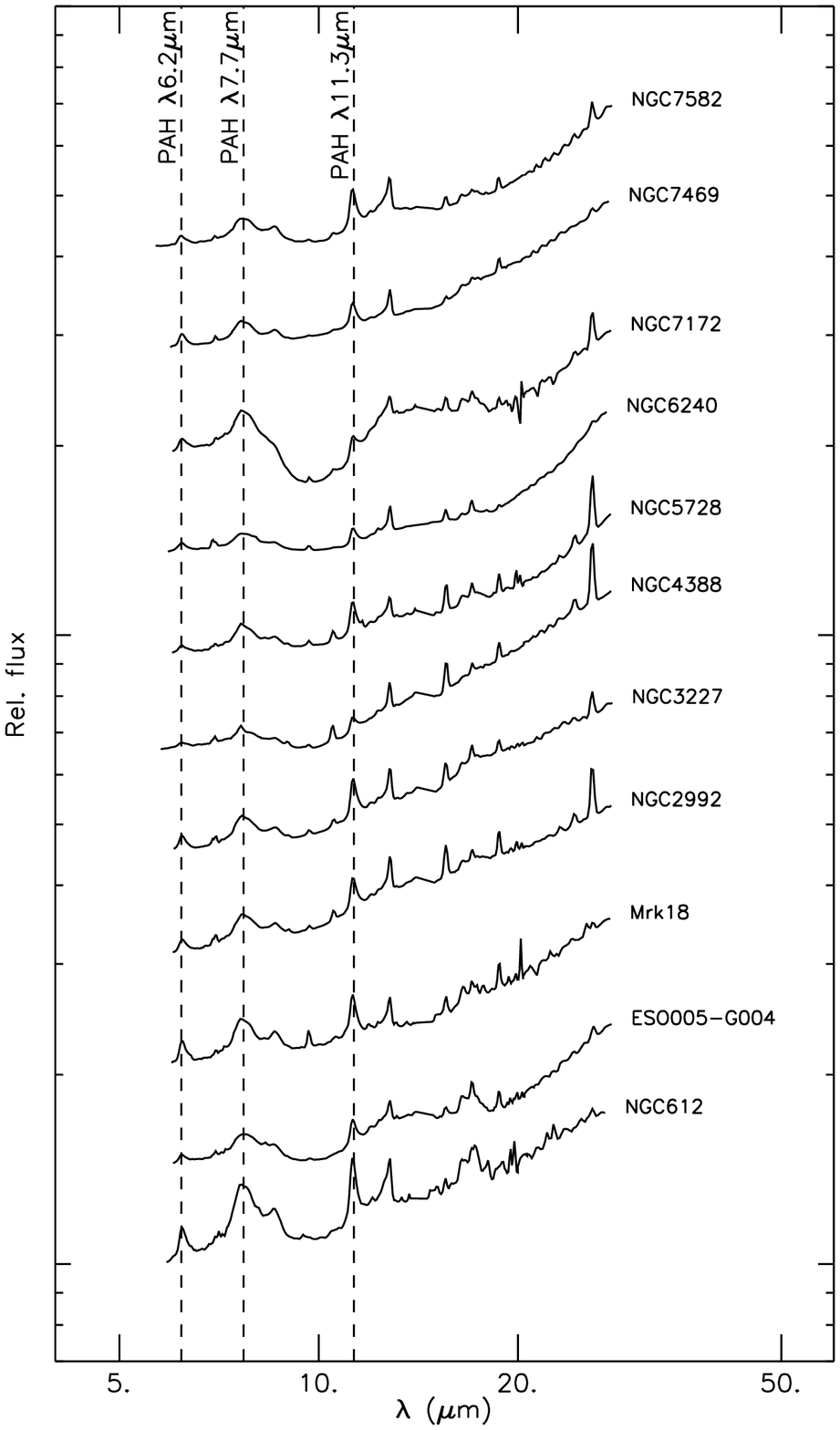}
\caption{IRS spectra of the 11 BAT/IRS AGNs that we classify as having
  SB-dominated IR SEDs on the basis of the presence of a prominent PAH
  feature at 6.2~\micron.  Also indicated are the PAH features at
  7.7~\micron\ and 11.4~\micron\, although these were not used to
  distinguish between AGN and SB dominated systems.}
\label{W09_Si}
\end{figure*}

To determine whether the \fratio\ flux ratio can distinguish between
SB and AGN dominated systems, we classified the BAT/IRS sample into
those objects with a) a prominent PAH feature at 6.2~\micron, which we
assume is a good indicator of a strong starburst component
(SB-dominated; e.g., \citealt{Genzel98}; 11 objects), and b) a
power-law MIR to FIR continuum with no sign of any PAH feature at
6.2~\micron, which we assume indicates an AGN-dominated object (25
objects). The 6.2~\micron\ PAH feature is preferred over other lines
as it lies in a region of the IR spectrum that is largely free of
other strong spectral features, in particular the silicate
absorption/emission band at 9.7~\micron. We show the {\it
  Spitzer}-IRS spectra of the BAT/IRS sample in Fig.~A1 (AGN-dominated
systems) and Fig.~A2 (SB-dominated systems).

\begin{figure*}
\includegraphics[angle=0,width=120mm]{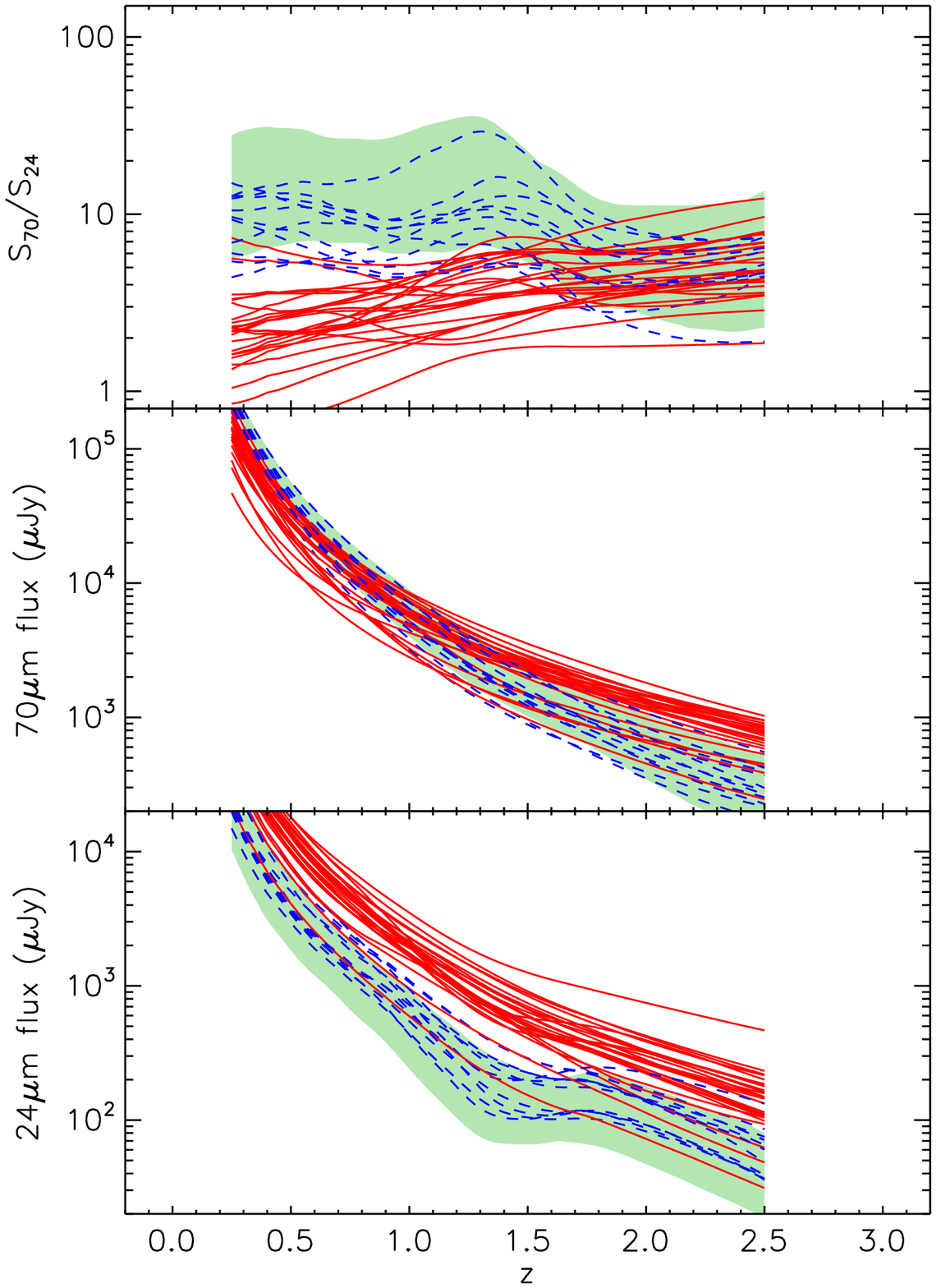}
\caption{\textit{Top}:The expected \fratio\ flux ratio of the AGNs in
  the BAT/IRS AGNs over the redshift range
  $z=0.25-2.5$. \textit{Middle}: The expected observed 70~\micron\
  flux densities of the BAT/IRS AGNs over the redshift range
  $z=0.25-2.5$, each normalised to \Lir$=10^{12}$~\Lsun.
  \textit{Bottom}: The expected observed 24~\micron\ flux densities of
  the BAT/IRS AGNs over the redshift range $z=0.25-2.5$, again
  normalised to \Lir$=10^{12}$~\Lsun.  We show the ratios and flux
  densities as expected if the AGNs were to be observed at
  $z=0.25-2.5$.  Solid, red lines refer to AGN-dominated and dashed
  blue lines to SB-dominated AGNs.  The green shaded areas indicate
  the range of ratios and flux densities spanned by the
  \protect\cite{Brandl06} starburst galaxies over the redshift range
  $z=0.25-2.5$.}
\label{W09_Tracks}
\end{figure*}

To calculate the expected \fratio\ flux ratio tracks each of for the
36 BAT/IRS AGNs as a function of redshift, we shift each spectrum by a
factor of $1+z$ in wavelength (where $z$ is increased from 0.25 to 2.5
in steps of 0.025) and pass the resulting spectrum through the MIPS
24~\micron\ and 70~\micron\ filters. To determine the observed
70~\micron\ fluxes at $z < 0.75$ requires knowledge of the IR SED
beyond the wavelength coverage of the {\it Spitzer}-IRS spectra
(i.e.,\ $>38$~\micron). To provide this longer wavelength coverage, we
extrapolated the longest wavelength bin of the IRS spectra to the
60~\micron\ and 100~\micron\ flux densities from the \textit{IRAS}
Point Source and Faint Source Catalogues\footnote{Available from:
  http://irsa.ipac.caltech.edu/applications/Gator/index.html}.
Thirty-two of the 36 AGNs with IRS data have well constrained \IRAS\
60~\micron\ and 100~\micron\ fluxes.  The remaining 4 (all classed as
AGN dominated) have \textit{IRAS} 60~\micron\ or 100~\micron\ flux
densities that are flagged as upper limits in the \IRAS\
catalogues. However, including these faint sources in our average AGN
SED by assuming the upper limits as detections has no effect on any of
our results.  In \Fig{W09_Tracks}, we show the \fratio\ flux ratios
derived from the BAT/IRS sample as a function of redshift. There is a
clear separation between AGN and SB dominated systems at $z \lesssim
1.5$, showing that the \fratio\ flux ratio can be used as an efficient
method to determine whether AGN or star-formation activity dominates
the IR emission of X-ray detected AGNs.

Calculating $L_{\rm IR}$ for high redshift sources using
the 24~\micron\ flux density alone is susceptible to systematic errors
caused by spectral features that are redshifted into this waveband at
$z>0.7$.  We can use the BAT/IRS sample to establish the uncertainties
in deriving \Lir\ from either 70~\micron\ or 24~\micron\ data.  In the
lower panels of \Fig{W09_Tracks} we show the expected observed frame
70~\micron\ and 24~\micron\ fluxes for the BAT/IRS sample, if observed
at z=0.25-2.5 and normalised to $L_{\rm IR}=10^{12}$~\Lsun\ (the
normalisation factor, $10^{12}$~\Lsun/$L_{\rm IR}$, is derived by
calculating $L_{\rm IR}$ from the \IRAS\ flux densities using the
equations presented in Table 1 of \citealt{Sanders96}).  Depending on
the shape of the IR SED, sources of the same $L_{\rm IR}$ can differ
in their observed 24~\micron\ flux density by up to a factor of
$\approx$12 at $z > 1$, compared to factor of only $\approx$3 at
70~\micron.  This range of 70~\micron\ flux density is almost
independent of redshift out to $z = 2.5$, indicating that
$S_{70}$ is up 4 times more accurate for determining $L_{\rm IR}$ when
no information on the SED shape is available.

\label{lastpage}
\end{document}